\documentclass[journal]{IEEEtran}
\usepackage{graphics} 
\usepackage{epsfig} 
\usepackage{mathptmx} 
\usepackage{times} 
\usepackage{amsmath} 
\usepackage{amssymb}  
\usepackage{upgreek}
\usepackage{latexsym,,graphicx,epsf,cite,url,bbm,float,subfig,amsbsy}
\usepackage{ifpdf}
\usepackage{epstopdf}
\usepackage{color}
\usepackage{tabularx}
\usepackage{stfloats}
\usepackage{tikz}
\usepackage[percent]{overpic}
\usepackage{arydshln}
\usepackage{multirow}

\usepackage{url}
\urldef{\mailsa}\path|{sayin2, basar1}@illinois.edu|

\makeatletter
\DeclareFontFamily{U}  {MnSymbolF}{}
\DeclareSymbolFont{symbolsMN}{U}{MnSymbolF}{m}{n}
\SetSymbolFont{symbolsMN}{bold}{U}{MnSymbolF}{b}{n}
\DeclareFontShape{U}{MnSymbolF}{m}{n}{
    <-6>  MnSymbolF5
   <6-7>  MnSymbolF6
   <7-8>  MnSymbolF7
   <8-9>  MnSymbolF8
   <9-10> MnSymbolF9
  <10-12> MnSymbolF10
  <12->   MnSymbolF12}{}
\DeclareFontShape{U}{MnSymbolF}{b}{n}{
    <-6>  MnSymbolF-Bold5
   <6-7>  MnSymbolF-Bold6
   <7-8>  MnSymbolF-Bold7
   <8-9>  MnSymbolF-Bold8
   <9-10> MnSymbolF-Bold9
  <10-12> MnSymbolF-Bold10
  <12->   MnSymbolF-Bold12}{}
\DeclareMathSymbol{\tbigtimes}{\mathop}{symbolsMN}{2}
\newcommand*{\bigtimes}{%
  \DOTSB
  \tbigtimes
  \slimits@
}
\makeatother

\DeclareMathOperator*{\argmin}{argmin}

\newcommand{\nn}{\nonumber}

\newcommand{\vv}{\mathcal{V}}

\newcommand{\E}{\mathbb{E}}
\newcommand{\N}{\mathbb{N}}
\newcommand{\A}{\textrm{A}}
\newcommand{\F}{\textrm{F}}
\newcommand{\C}{\textrm{C}}
\newcommand{\rmS}{\textrm{S}}
\newcommand{\T}{\textrm{T}}
\newcommand{\LL}{{\mathcal L}}

\newcommand{\rx}{\pmb{x}}

\newcommand{\rw}{\pmb{w}}
\newcommand{\rv}{\pmb{v}}
\newcommand{\ry}{\pmb{y}}

\newcommand{\ru}{\pmb{u}}
\newcommand{\rtheta}{\pmb{\theta}}

\newcommand{\bx}{\,\pmb{\bar{x}}}

\newcommand{\ou}{\pmb{y}}

\newcommand{\hx}{\,\pmb{\hat{x}}}

\newcommand{\rtx}{\pmb{\check{x}}}

\newcommand{\rs}{\pmb{s}}

\newcommand{\setS}{\Upsilon}
\newcommand{\setC}{\Gamma}
\newcommand{\tr}{\mathrm{tr}}
\newcommand{\deltau}{\,\delta\pmb{u}\,}

\newcommand{\tQ}{\check{Q}}

\newcommand{\bA}{\bar{A}}
\newcommand{\bB}{\bar{B}}
\newcommand{\bQ}{\bar{Q}}

\newcommand{\uz}{\underline{z}}
\newcommand{\Ek}{\mathbb{E}}

\newenvironment{psmallmatrix}
  {\left[\begin{smallmatrix}}
  {\end{smallmatrix}\right]}

\begin{document}

\title{Secure Sensor Design Against Undetected Infiltration: Minimum Impact-Minimum Damage}

\author{Muhammed~O.~Sayin and~Tamer~Ba\c{s}ar,~\IEEEmembership{Life~Fellow,~IEEE}
\thanks{This research was supported by the U.S. Office of Naval Research (ONR) MURI grant N00014-16-1-2710. The authors are with the Department of Electrical and Computer Engineering, University of Illinois at Urbana-Champaign, Urbana, IL 61801 USA. E-mail: \{sayin2,basar1\}@illinois.edu}}

\maketitle

\begin{abstract}
We propose a new defense mechanism against undetected infiltration into controllers in cyber-physical systems. To this end, we cautiously design the outputs of the sensors that monitor the state of the system. Different from the defense mechanisms that seek to detect infiltration, the proposed approach seeks to minimize the damage of possible attacks before they have been detected. Controller of a cyber-physical system could have been infiltrated into by an undetected attacker at any time of the operation. Disregarding such a possibility and disclosing system's state without caution benefits the attacker in his/her malicious objective. Therefore, secure sensor design can improve the security of cyber-physical systems further when incorporated along with other defense mechanisms. We, specifically, consider a controlled Gauss-Markov process, where the controller could have been infiltrated into at any time within the system's operation. In the sense of game-theoretic hierarchical equilibrium, we provide a semi-definite programming based algorithm to compute the optimal linear secure sensor outputs and analyze the performance for various scenarios numerically.
\end{abstract}


\section{Introduction}
\IEEEPARstart{C}{yber-physical} systems, incorporating both physical and cyber parts, e.g., process control systems, robotics, smart grid, and autonomous vehicles, have resulted in new and distinct challenges for control system design, e.g., specifically, security-related challenges due to cyber attacks \cite{giraldo17,humayed17}. Different from external random disturbances, cyber attacks can be very target specific and persistent by attacking stealthily for long term benefits. Recently in 2014, cyber-physical, e.g., process control, systems in energy and pharmaceutical industries have been infiltrated into by Dragonfly Malware, which intervened in the systems over a long period of time without being detected \cite{nelson16}. However, isolation from the cyber networks are also not perfectly sufficient any more. In 2010, StuxNet Worm caused substantial damage on certain ``isolated" supervisory control and data acquisition (SCADA) systems \cite{StuxNet}. Therefore, developing novel formal security mechanisms against advanced and persistent threats that can cause substantial damage without being detected plays a vital role in the security of cyber-physical systems.

Beyond exploiting uncertainties in the systems, e.g., due to random disturbances, advanced and persistent attackers can also seek to deceive the detection mechanisms by manipulating monitoring signals used by the detectors. For example, in \cite{ref:liu09}, the authors have introduced false data injection attacks, where the attackers can inject data into the sensor outputs, in the context of state estimation, and characterized undetectable attacks. Based on the deceptive attacker model in \cite{ref:liu09}, the existing studies mainly focus on characterizing the vulnerabilities of control systems against such undetectable attacks (which can deceive the detectors) and designing counter measures to be able to detect them. 

In discrete-time linear-quadratic-Gaussian (LQG) systems, in \cite{ref:mo09}, the authors have introduced replay attacks and proposed a defense mechanism against such attacks. Replay attacks take place during the steady state of the system, and the attacker records and replays the sensor outputs so that the detectors using those signals cannot detect any anomalies. Note that the signals are expected to be similar at steady state. As a defense mechanism, the authors have proposed to inject an independent signal into the control input to detect such attacks in the expense of degraded control performance. An optimal defense strategy with respect to the probability of detection has been formulated in \cite{ref:mo14}. Again in LQG systems, in \cite{ref:mo12}, the same set of authors have introduced integrity attacks, where the attacker can inject data into sensor outputs and control inputs, and characterized the reachable set that the attacker can drive the system to without being detected (via innovation based failure detectors \cite{ref:willsky76}). They have also provided necessary conditions for unbounded reachable set, i.e., conditions where the attacker can destabilize the system. 

Within deterministic control scenarios, in \cite{ref:teixeira12}, the authors have analyzed zero-dynamics attacks that do not depend on online information, i.e., open-loop stealthy attacks, and provided an algorithm to reveal all such attacks by adding new measurements, similar to \cite{ref:mo09}. Again within deterministic control scenarios, in \cite{ref:pasqualetti13}, the authors have provided a unified framework for false data injection and replay attacks, and formulated the limitations of monitoring-based detection mechanisms. In \cite{ref:fawzi14}, the authors have analyzed tolerance of control systems to false data injection attacks on a subset of sensors in the deterministic settings and proposed decoding schemes to estimate the state via corrupted measurements. They have also introduced a secure control loop that can enhance the decoding performance and, correspondingly, the resilience of the system.  

The attackers can also have adversarial control objectives. In \cite{ref:chen16,ref:chenACC16}, the authors have analyzed such attacks, where the attacker both seeks to be undetected and drive the state of the system according to his/her adversarial goal by manipulating both sensor outputs and control inputs together. Recently, \cite{ref:zhang17} has analyzed optimal attack strategies to maximize the quadratic cost of a system with linear Gaussian dynamics without being detected, where the stealthiness is measured in terms of the Kullback-Leibler distance between the realized and the desired state behaviors. In the optimal attack, the attacker injects independent Gaussian noise having certain variance into the control input. In another recent study \cite{ref:miao17}, the authors have proposed linear encoding schemes for sensor outputs of an LQG system in order to enhance detectability of false data injection attacks while the coding matrix is assumed to be unknown by the attackers, which can be mitigated via time-varying coding matrices. In spite of these extensive studies, we still have significant and yet unexplored problems about how to enhance security against undetected attacks, i.e., impact of attacks before detection.

In this paper, we address primarily the following two questions: ``If we have already designed the sensor outputs, to what extent would we have secured the system against undetected infiltration into the controllers?" Further, ``what would be the best linear sensor outputs that can lead to both minimum impact and minimum damage on the ordinary operations of the systems?" The damage due to inconspicuous (undetectable, or difficult to detect) attacks with long term control objectives is our main concern in this paper. We can classify such attacks as ``advanced and persistent threats", since they are advanced by being very target specific and persistent by being inconspicuous. Therefore, we propose to be cautious while disclosing the state information to the controller due to the possibility of undetected infiltration. However, as a system designer, we should not take precautions as if the cyber part of the system is compromised due to just a possibility, since that would impact the ordinary operations of the system substantially. Combining these seemingly opposing goals all together, we seek to design sensor outputs cautiously with {\em minimum impact} and {\em minimum damage} on the system's operations.

To obtain explicit results, we specifically consider systems with linear Gaussian dynamics and quadratic control objectives, which have various applications in industry \cite{ref:zhang17} from manufacturing processes to aerospace control. We consider the possibility of infiltration into the controller of the system by various attackers at any stage within the time horizon. The attackers have long term control objectives and attack stealthily. To this end, they include soft constraints on the energy of the state and the deviation of the constructed control inputs from system-desired ones, which would have been constructed if the attacker rather had a friendly objective. Such constraints are, especially, against the detection mechanisms that take actions when such deviations exceed certain thresholds. 

We note that the sensors could also be infiltrated into by the attackers, which can cancel the proposed approach via a shortcut to the state if the sensors have access to the state realizations. To mitigate that, we consider the scenarios where the sensors do not have access to the actual state realizations. All the sensor strategies, defining the relation between the state and the sensor output, are selected beforehand to minimize the expected loss and fixed (can be time-variant, yet not controlled) during the operation. Therefore, we can design the sensor outputs off-line, i.e., in advance, which leads to a hierarchical structure between the sensor and the controller of the system (even when he/she is adversarial).

Due to the stochastic nature of the problem, i.e., due to the state noise, any open-loop control strategy of an attacker could not drive the system in his/her desired path effectively \cite{kumarStochasticBook}. Therefore, regardless of whether the controller has an adversarial objective or not, he/she needs to construct a closed-loop control input based on the designed sensor outputs while knowing the relationship between the sensor output and the state. This implies that the interaction between the sensor and the controller of the system could be analyzed as a game-theoretic hierarchical equilibrium, where the sensor leads the game by announcing his/her strategies beforehand. Therefore, while designing the sensor outputs, we should consider both adversarial and friendly control outputs and the possibility of infiltration over the time horizon. 

Particularly, we seek to formulate the best linear sensor strategies for controlled Gauss-Markov processes. We consider a different time scale for the infiltration into the system and we formulate the optimal sensor strategies in a Bayesian setting based on given infiltration statistics.  Note that since the sensor strategies are set to be linear, the problem is an LQG control problem and correspondingly the optimal, friendly, control policy is linear in the conditional estimate of the state given all the sensor outputs. We first compute the best control inputs of friendly and adversarial controllers for any given linear sensor strategies and any time when they become in charge of the controller. Corresponding to these optimal strategies, we provide a semi-definite programming (SDP) based algorithm to design the optimal (memoryless) linear secure sensor strategies. Furthermore, we analyze the sensitivity of the design against inaccurate perception of the underlying statistics numerically. We note that in \cite{sayinGameSec17}, we have introduced secure sensor design against advanced and persistent threats in cyber-physical systems, but have not completely solved the problem. Here, we consider different attack models that have soft constraints on the energy of the state and the deviation of the constructed control inputs from the system-desired ones, and we consider more comprehensive scenarios, where the controller can be infiltrated into or an adversarial infiltration could be detected at any time within the time horizon. 

In the design of secure cyber-physical systems, each additional security layer leads to new monetary and computational costs \cite{brangetto15}. In particular, there are fundamental trade-offs in terms of investment on the security mechanisms and the value of the protected assets or securing the system and maintaining the ordinary operations. In order to offer better trade-offs, we aim to propose a defense mechanism that does not require any additional online computational load with minimum impact and minimum damage on the ordinary operations of the system. To summarize, we can list the main contributions of this paper as follows:
\begin{itemize}
\item We introduce secure sensor design against various inconspicuous attackers with control objectives, which can infiltrate into the controller of a cyber-physical system at any time during the operation.
\item Given any linear sensor strategies and the underlying linear quadratic Gaussian dynamics, we compute the optimal attack strategies depending on the infiltration time.
\item We provide a practical algorithm to compute the optimal linear memoryless sensor strategies in the sense of game-theoretic hierarchical equilibrium.
\item We also analyze sensitivity of the proposed algorithm against inaccurate perception of attack statistics.
\end{itemize}

The paper is organized as follows: In Section \ref{sec:prob}, we provide the secure sensor design framework. In Section \ref{sec:game}, we formulate the associated multi-stage Bayesian Stackelberg game. In Section \ref{sec:controller}, we characterize the optimal controller response strategies for given sensor strategies. We compute the corresponding optimal sensor strategies in Section \ref{sec:sensor}. In Section \ref{sec:examples}, we examine the performance of the proposed scheme under various scenarios numerically. We conclude the paper in Section \ref{sec:conclusion} with several remarks and possible research directions. 

\noindent{\bf Notations:}  For an ordered set of parameters, e.g., $x_1,\cdots,x_n$, we define $x_{[k,l]} := x_k,\cdots,x_l$, where $1\leq k \leq l \leq n$. ${\mathbb N}(0,.)$ denotes the multivariate Gaussian distribution with zero mean and designated covariance. We denote random variables by bold lower case letters, e.g., $\rx$. For a random variable $\rx$, $\hx$ is another random variable corresponding to its posterior belief conditioned on certain other random variables that will be apparent from the context. For a vector $x$ and a matrix $A$, $x'$ and $A'$ denote their transposes, and $\|x\|$ denotes the Euclidean ($L^2$) norm of the vector $x$. For a matrix $A$, $\mathrm{tr}\{A\}$ denotes its trace. We denote the identity and zero matrices with the associated dimensions by $I$ and $O$, respectively, while ${\bf 1}$ (or ${\bf 0}$) denotes a vector whose entries are all $1$ (or $0$). For positive semi-definite matrices $A$ and $B$, $A\succeq B$ means that $A-B$ is also a positive semi-definite matrix. $A\otimes B$ denotes the Kronecker product of the matrices $A$ and $B$.

\section{Problem Formulation}\label{sec:prob}
Consider a controlled stochastic system described by the following equations:
\begin{align}
&\rx_{k+1} =A\,\rx_k + B\,\ru_k + \rv_k,\label{eq:state}\\
&\ou_{k}=D\,\ru_k + \rw_k, \label{eq:observedControl}
\end{align}
for $ k = 1,2,\ldots,n$, where\footnote{Even though we consider time invariant matrices $A, B$, and $D$ for notational simplicity, the provided results could be extended to time-variant cases rather routinely. Furthermore, we consider all the random parameters to have zero mean; however, the derivations can be extended to non-zero mean case in a straight-forward way.} $A\in\mathbb{R}^{m\times m}$, $B\in\mathbb{R}^{m\times r}$, $D\in\mathbb{R}^{r\times r}$, and $\rx_k\sim\N(0,\Sigma_k)$, $k=1,\ldots,n$. The additive state and control input noise sequences $\{\rv_k\}$ and $\{\rw_k\}$ are white Gaussian vector processes, i.e., $\rv_k \sim \N(0,\Sigma_v)$ and $\rw_k \sim \N(0,\Sigma_w)$; and are independent of the initial state $\rx_1$ and of each other. We assume that the matrix $A$ is non-singular, and the auto-covariance matrices $\Sigma_1$ and $\Sigma_v$ are positive definite while $\Sigma_w$ is positive semi-definite. The closed loop control vector $\ru_k \in \mathbb{R}^r$ is given by
\begin{equation}\label{eq:control}
\ru_k = \gamma_k(\rs_{[1,k]}),
\end{equation}
where $\gamma_k(\cdot)$ can be any Borel measurable function from $\mathbb{R}^{mk}$ to $\mathbb{R}^r$. The sensor output $\rs_k\in\mathbb{R}^m$ is given by
\begin{equation}\label{eq:output}
\rs_k = \eta_k(\rx_k),
\end{equation}
where $\eta_k(\cdot)$ can be any {\em linear} function from $\mathbb{R}^{m}$ to $\mathbb{R}^m$. And $\ou_k \in \mathbb{R}^r$, given by \eqref{eq:observedControl}, denotes the noisy observation of the control input $\ru_k$. 

\begin{figure}[t!]
  \centering
  \includegraphics[width=.48\textwidth]{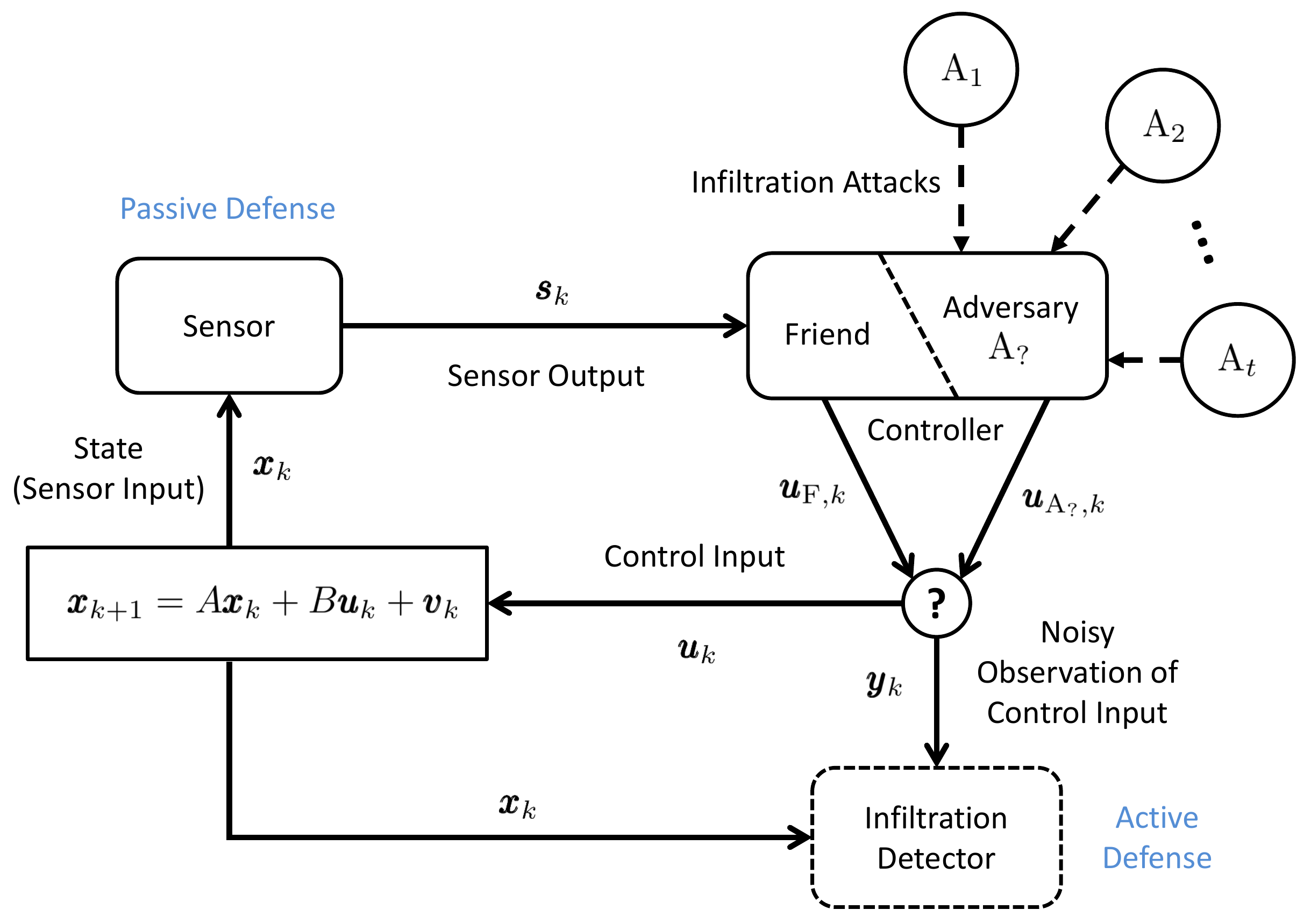}\\
  \caption{Cyber physical system including a sensor and a controller.}\label{fig:model}
\end{figure}

We have two separate agents: Sensor (S) and Controller (C). At each stage $k=1,\ldots,n$, the agents construct $\rs_k$ and $\ru_k$ according to their own objectives. In particular, S chooses $\eta_k(\cdot)$ from the strategy space $\setS$, which, for each $k$, is the set of all linear functions from $\mathbb{R}^{m}$ to $\mathbb{R}^m$, i.e., $\eta_k\in\setS$ and $\rs_k = \eta_k(\rx_{k})$. This implies that for each $\eta_k\in\setS$, there exists a matrix $\LL_k \in \mathbb{R}^{m\times m}$ such that
\begin{equation}\label{eq:LL}
\rs_k = \LL_k'\rx_k,
\end{equation} 
almost surely on $\mathbb{R}^m$. C chooses $\gamma_k(\cdot)$ from the strategy space $\setC_k$, which is the set of all Borel measurable functions from $\mathbb{R}^{mk}$ to $\mathbb{R}^r$, i.e., $\gamma_k \in \setC_k$ and $\ru_k = \gamma_k(\rs_{[1,k]})$. 

{\em Sensor and Controller Objective.}
As in a stochastic control scenario \cite{kumarStochasticBook}, S and C can have a common finite horizon\footnote{E.g., horizon length is $n$.} quadratic cost function:
\begin{align}
J_{\F}(\eta_{[1,n]};\gamma_{[1,n]}) = \E\left\{\sum_{k=1}^{n}\|\rx_{k+1}\|_{Q_{\F}}^2 + \|\ru_k\|_{R_{\F}}^2\right\},\label{eq:loss}
\end{align}
where\footnote{For notational simplicity, we consider time-invariant $Q_{\F}$ and $R_{\F}$. However, the provided results could be extended to time-variant cases rather routinely.} $Q_{\F}\in\mathbb{R}^{m\times m}$ is positive semi-definite and $R_{\F} \in \mathbb{R}^{r\times r}$ is positive definite. Note that $\ru_k = \gamma_k(\rs_{[1,k]})$ while $\rs_k = \eta_k(\rx_{k})$ almost surely. Correspondingly, S could disclose the state $\rx_k$ directly so that C could drive the state in their commonly desired path \cite{optControl, kumarStochasticBook}. However, in a cyber physical system, the system is vulnerable to adversarial infiltration attacks that seek to drive the state of the system away from the system's desired target as seen in Fig. \ref{fig:model}. We call such attacks ``advanced persistent threats", which are advanced by being very target specific, i.e., the attacker knows the underlying state recursion, and persistent by avoiding infiltration detection. Therefore, S, i.e., the sensor designer, should anticipate the likelihood of adversarial infiltration into C, i.e., the possibility that C can be an adversary, and select $\eta_k\in\setS$ accordingly. 

\noindent
{\bf Remark 1.} 
{\em We note that the sensor output $\rs_k$ only depends on the current state $\rx_k$, i.e., $\eta_k$ is memoryless. Otherwise, \textup{S} would need to have access to the state information in order to store and  to be able to use them in the future stages. However, similar to \textup{C}, \textup{S} can also be infiltrated into by the attackers, which would neutralize \textup{S}'s effort to design sensor outputs strategically via a shortcut to the state information.} 

\noindent
{\bf Remark 2.} {\em In control system design, sensors are designed and implemented in advance, and system engineers design the controllers knowing the relation between the sensor output and the underlying state. Correspondingly, an attacker that has infiltrated into the system can be aware of how the sensor outputs have been constructed and can design his/her attack accordingly. Therefore, there exists a hierarchy between \textup{S} and \textup{C} such that \textup{C} can have access to \textup{S}'s strategies.}

{\em Infiltration Detection.}
We note that if the control inputs could have been monitored perfectly, then any deviation of the control input from the system-desired one could have been detected instantly since both sensor outputs and control inputs will be accessible. Therefore, in this paper, we address the scenarios where the control input cannot be monitored perfectly. As an example of such scenarios, the infiltration detection mechanism can have access to noisy control input observation $\ou_k \in \mathbb{R}^r$ and the state $\rx_k$ as seen in Fig. \ref{fig:model}. In the scope of this work, we will not consider the details of how the infiltration detector operates except that the advanced and persistent attackers are aware of the presence of an infiltration detector that can have access to the state (correspondingly the sensor outputs) and noisy versions of the control inputs.

{\em Inconspicuous Infiltration Attacks into \textup{C}.}
As seen in Fig. \ref{fig:model}, C is under infiltration attacks by the (advanced) attackers $\A_1,\ldots,\A_t$ over the time horizon $k=1,\ldots,n$ and such attacks may be successful or not in infiltrating into C. As mentioned earlier, being advance refers to being target specific with knowledge about underlying system dynamics while also avoiding detection mechanisms by attacking inconspicuously. Therefore, C can be a friend or an adversary within the time horizon while S may not know C's type surely until an infiltration detection takes place, which may be less likely due to inconspicuousness of the attacks. C observes $\rs_{[1,k]}$, knows S's strategies $\eta_{[1,k]}$ due to a hierarchy between the agents, and, via a strategy $\gamma_k \in \setC_k$, can construct a closed-loop control input $\ru_k$, yet the state and the control input can be monitored by the infiltration detector. Therefore, as an attack model, we consider the situation where the attacker $\A_i$, $i=1,\ldots,t$, selects $\gamma_{\A_i,k}\in\setC_k$, $k=\kappa,\ldots,n$, where $\kappa$ denotes the infiltration time, to minimize the cost function:
\begin{align}
J_{\A_i,\kappa}(\eta_{[1,n]};\cdot,\gamma_{\A_i,[\kappa,n]}&) = \E\bigg\{\sum_{k=\kappa}^{n} \|\rx_{\A_i,k+1}-z_i\|_{Q_{\A_i}}^2\nn\\
 +&\lambda_i\|\rx_{\A_i,k+1}\|_{Q_{\F}}^2 + \|\ru_{\A_i,k} - \ru_{\F,k}\|_{R_{\A_i}}^2\bigg\}, \label{eq:lossA}
\end{align}
where $\lambda_i\geq 0$, ``$\cdot$" as an argument of the cost function \eqref{eq:lossA} refers to the C's strategies $\gamma_1,\ldots,\gamma_{\kappa-1}$, which are not selected by $\A_i$; and $z_i\in\mathbb{R}^m$ is the desired state that the adversary seeks to drive the system to. The matrices\footnote{For notational simplicity, we consider time-invariant $\lambda_i$, $Q_{\A_i},R_{\A_i}$.} $Q_{\A_i}\in\mathbb{R}^{m\times m}$ are positive semi-definite, and $R_{\A_i}\in \mathbb{R}^{r\times r}$ are positive definite. Here, $\rx_{\A_i,k}$ denotes the state driven by the adversarial control input $\ru_{\A_i,k}$; while $\ru_{\F,k}$ denotes the control input that would have been constructed if C was a friend.

Particularly, the last two terms in \eqref{eq:lossA} are soft constraints to avoid infiltration detection by being close to the expected behavior of the system, e.g., small energy of the state $\rx_{\A_i,k+1}$, and small deviations of $\ru_{\A_i,k}$ from $\ru_{\F,k}$. Note that $\lambda_i\geq 0$ can also be zero. We also note that deviation of the state $\rx_{\A_i,k+1}$ from the system-desired $\rx_{\F,k+1}$ is equivalent to
\[
\rx_{\A_i,k+1}-\rx_{\F,k+1} = B(\ru_{\A_i,k}-\ru_{\F,k}).
\]
Furthermore, deviation of the observed control input\footnote{By depending on $\ru_{\A_i,k}$, observed control inputs also depend on $\A_i$'s actions. Therefore, we show this dependence explicitly by $\ry_{\A_i,k}:=\ry_k$. Similarly, the system-desired observation, i.e., $\F$ would have been in charge of C, is denoted by $\ry_{\F,k}$.} $\ry_{\A_i,k}$  from the system desired $\ry_{\F,k}$ is equivalent to
\[
\ry_{\A_i,k+1}-\ry_{\F,k+1} = D(\ru_{\A_i,k}-\ru_{\F,k}).
\]
Therefore, via $R_{\A_i} \in \mathbb{R}^{r\times r}$, the attacker can take precautions against the thresholding-based detection mechanisms that check the deviation of the state or the control input from the system-desired ones. We also note that $\ru_{\F}$ would have been constructed by C if there were no infiltration. Correspondingly, the attacker can construct $\ru_{\A_i,k}$ by adding another signal on top of $\ru_{\F,k}$ such that $\ru_{\A_i,k} = \ru_{\F,k} + \deltau_{\A_i,k}$ as in  \cite{ref:chen16,ref:chenACC16}. Therefore, the last term in \eqref{eq:lossA} also corresponds to a soft energy constraint on $\deltau_{\A_i,k}$.

\begin{figure}
  \centering
  \includegraphics[width=.35\textwidth]{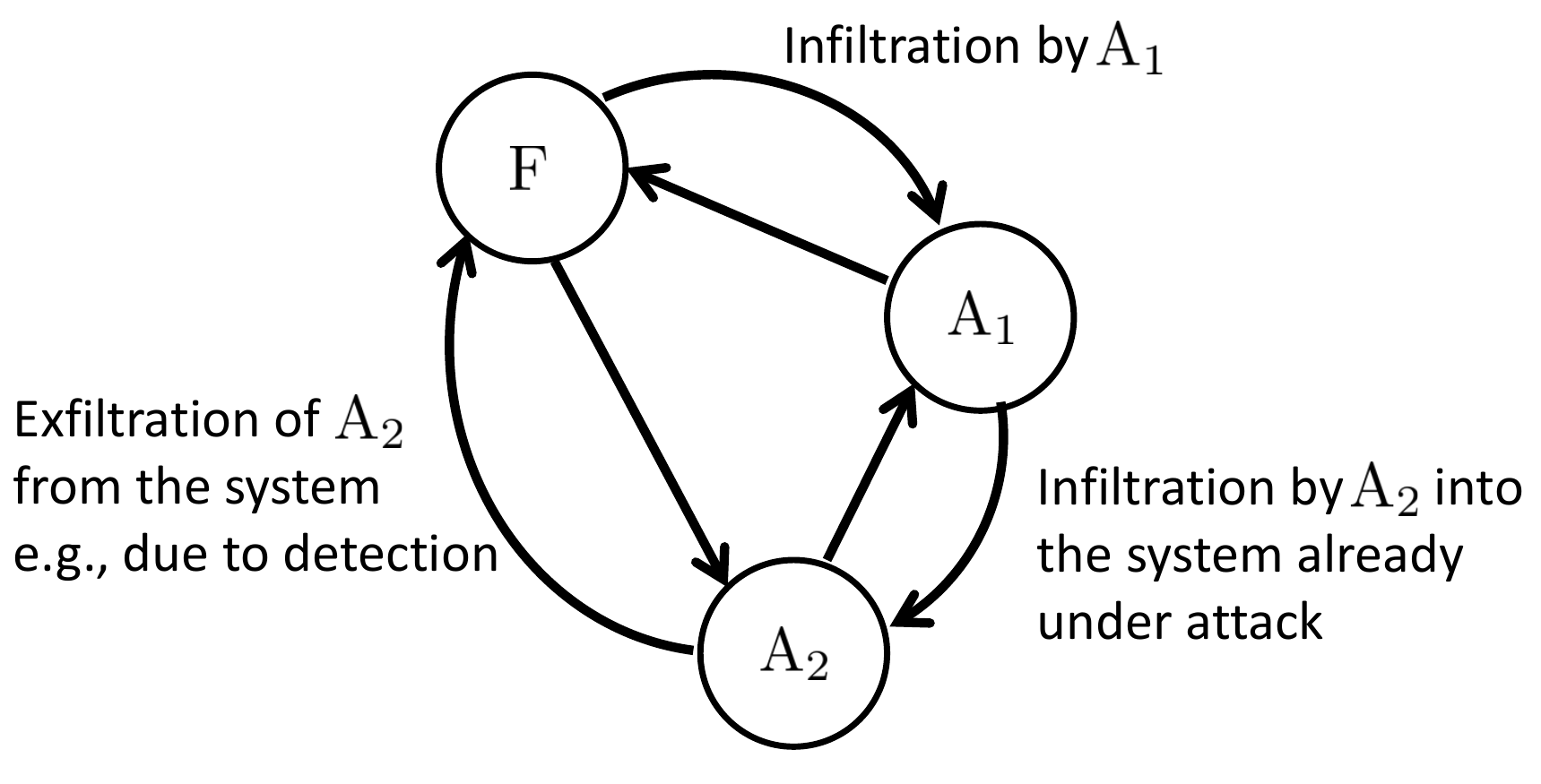}
  \caption{Possible transitions among the agents $\F,\,\A_1,$ and $\A_2$ to be in charge of C within the time horizon.}\label{fig:switch}
\end{figure}

In the following section, we provide a game theoretical formulation to analyze the interaction between the agents.

\section{Hierarchical Equilibrium Formulation} \label{sec:game}

Cyber-physical systems are vulnerable to infiltration attacks. Within the time horizon various attackers can infiltrate into the system as well as defense mechanism can detect infiltration and take appropriate actions accordingly. As an example, Fig. \ref{fig:switch} demonstrates possible transitions among the agents, $\F$, $\A_1$, and $\A_2$, to be in charge of C. When an attacker, e.g., $\A_1$, infiltrates into the system, $\A_1$ becomes in charge of C and can construct the control input according to his/her adversarial objective. Furthermore, when the attackers are not communicating with each other, an attacker, e.g., $\A_2$, may not know whether the system is already under attack or not and correspondingly may infiltrate into the system while $\A_1$ is still in charge of C. Furthermore, there can be infiltration detection due to the active defense mechanisms monitoring the control and sensor inputs as seen in Fig. \ref{fig:model}. 

\noindent
{\bf Remark 3.}
{\em We point out that when infiltration has been detected, the system could prefer to use different sensor outputs since the uncertainty about \textup{C} has been removed. As an example, if the attacker cannot be forced to exfiltrate from the system immediately, the system can prefer to use sensor outputs designed specifically against the adversarial objectives of the attacker. Or if \textup{F} becomes in charge of \textup{C} after detection, as long as the system ensures \textup{F} is in charge, direct state disclosure can be preferred. Therefore, we consider scenarios where the system switches his/her operation mode and uses different sensor outputs, once an infiltration has been detected.}

In order to model uncertainty of the transitions between the agents to be in charge of C explicitly, we consider a jump process $\{\rtheta_j \in \Theta\}$, where $\Theta:=\{\F,\A_1,\ldots,\A_t,\T\}$ and transitions can occur on a different time scale called transition time, e.g., $\kappa\in\{\delta, 2\delta,\ldots,N\delta\}$, where $\delta \in \mathbb{Z}$ and $N:=\lceil n/\delta\rceil$. The state $\T$ denotes the switch to a different operation mode, e.g., due to infiltration detection. Therefore, when the process jumps to the state $\T$, the horizon practically terminates for S. Correspondingly, for a given sequence of the jump process: $\theta_1,\ldots,\theta_{N}$, if there is a jump to $\T$, we let $N_T$ denote the index of the last state before the jump, and otherwise $N_T = N+1$, and let $\hbar(\theta_{[1,N]})$ be an ordered set including the non-terminating transition times, and $1$ and $\min\{N_T\delta,n+1\}$. As an illustrative example, let $n=100$ and $\delta = 30$ and consider the situation where $\F$ is in charge of C initially while $\A_1$ infiltrates into C at $k = 60$ and becomes in charge until detection at $k=90$. Therefore, we have $\hbar(\theta_{[1,N]}) = \{1,60,90\}$, which implies that while $\theta_1=\theta_2=\F$ is in charge during the interval $[1,60)$, $\theta_3=\A_1$ becomes in charge during the interval $[60,90)$ and the system switches its mode at $k=90$.

The underlying state recursion is common knowledge to both S and C (even if C can be an adversary). The type of C and, if C is an adversary, his/her objective are not known by S. However, S knows the statistics of the jump process $\{\rtheta_j\}$. As also noted in Remark 1, there is also a hierarchy \cite{basarDynamicBook} between the agents in the announcement of the strategies such that S leads the game by announcing and sticking to his/her strategies in advance, i.e., C knows $\eta_{[1,n]}$ in advance. Therefore, we can model such a scheme as a multi-stage Bayesian {\em Stackelberg} game, in which S is the leader.

\noindent
{\bf Remark 4.} {\em When the infiltration detector has access to the sensor outputs $\rs_k$ but not the states $\rx_k$, even though the attacker can also inject false data into the sensor outputs in order to avoid detection as in integrity attacks, e.g., \cite{ref:mo12,ref:chen16,ref:chenACC16}, due to the ``stochastic" state recursion \eqref{eq:state}, the attacker still needs the actual sensor outputs, which are designed by the system designer in advance. Therefore, secure sensor design framework can also play a crucial role for the security of the systems against integrity attacks.}

The agents S and C aim to minimize their cost functions by choosing the strategies $\eta_{[1,n]}$ and $\gamma_{[1,n]}$ while each strategy implicitly depends on the other. Due to the hierarchy, C's strategies $\gamma_{\theta,k} \in \setC_k$, $\theta\in\{\F,\A_1,\ldots,\A_t\}$, depending on his/her type, can also depend on S's strategies $\eta_{[1,k]}$ and when the agent becomes in charge of C. In order to show these dependences explicitly, henceforth, we denote C's strategies by $\gamma_{\theta,k}^{(\kappa)}(\eta_{[1,k]})$, which implies $\gamma_{\theta,k}^{(\kappa)}(\eta_{[1,k]})(\rs_{[1,k]}):=\gamma_{\theta,k}(\rs_{[1,k]})$. Then, for given S strategies $\eta_{[1,n]}$, we let $\Pi_{\theta,\kappa}(\eta_{[1,n]}) \subset\bigtimes_{k=\kappa}^n \Gamma_k$ be the reaction set of the agent $\theta\in\{\F,\A_1,\ldots,\A_t\}$ who becomes in charge of C at $\kappa$. And these reaction sets are given by:
\begin{align}
&\Pi_{\theta,\kappa}(\eta_{[1,n]}) := \argmin_{\substack{\gamma_{\theta,k}^{(\kappa)} \in \setC_k\\k=\kappa,\ldots,n}}\; J_{\theta,\kappa}\left(\eta_{[1,n]};\cdot,\gamma_{\theta,[\kappa,n]}^{(\kappa)}(\eta_{[1,n]})\right),\nn
\end{align}
where $\gamma_{\theta,[\kappa,n]}^{(\kappa)}(\eta_{[1,n]}):=\left\{\gamma_{\theta,\kappa}^{(\kappa)}(\eta_{[1,\kappa]}),\ldots,\gamma_{\theta,n}^{(\kappa)}(\eta_{[1,n]})\right\}$.
Due to the positive definiteness assumptions on $R_{\theta}$, in the following section, we will show that for each $\theta \in \{\F,\A_1,\ldots,\A_t\}$, the corresponding reaction set $\Pi_{\theta,\kappa}$ is an equivalence class such that all $\gamma_{\theta,[\kappa,n]}^{(\kappa)}\in\Pi_{\theta,\kappa}$ lead to the same random variable $\ru_{\theta,k}^*$ almost everywhere on $\mathbb{R}^r$. Therefore, the pair of strategies:
\begin{equation}\label{eq:pair}
\left[\eta_{[1,n]}^*;\left(\gamma_{\theta,[\kappa,n]}^{(\kappa)*}, \theta\in\Theta,\kappa\in\hbar\right)\right]
\end{equation}
attains the Stackelberg equilibrium provided that
\begin{subequations}\label{eq:equilibrium0}
\begin{align}
&\eta^*_{[1,n]} = \argmin_{\substack{\eta_{k}\in\setS,\\ k=1,\ldots,n}}\; \E\left\{\sum_{\kappa \in \hbar(\rtheta_{[1,N]})} \hspace{-.1in}J_{F,\kappa}^{\kappa_+}\left(\eta_{[1,n]};\cdot,\gamma_{\rtheta_{j},[\kappa,n]}^{(\kappa)*}(\eta_{[1,n]})\right)  \right\}\label{eq:all0}\\
&\gamma_{\theta,[\kappa,n]}^{(\kappa)*}(\eta_{[1,n]}) = \argmin_{\substack{\gamma_{\theta,k}^{(\kappa)}\in\setC_k,\\ k=\kappa,\ldots,n}} \; J_{\theta,\kappa}\left(\eta_{[1,n]};\cdot,\gamma_{\theta,[\kappa,n]}^{(\kappa)}(\eta_{[1,n]})\right),
\end{align}
\end{subequations}
where the expectation is taken over $\{\rtheta_{j}\}$, $\kappa_+$ is the next transition time after $\kappa$ in $\hbar(\theta_{[1,N]})$, $\rtheta_j$ refers to the agent that will be in charge of C in the interval $[\kappa,\kappa_+)$, and we define
\begin{align}
J_{F,\kappa}^{\kappa_+}\left(\eta_{[1,n]},;\cdot,\gamma_{\rtheta_{j},[\kappa,n]}^{(\kappa)*}(\eta_{[1,n]})\right) := J_{F,\kappa}\left(\eta_{[1,n]};\cdot,\gamma_{\rtheta_{j},[\kappa,n]}^{(\kappa)*}(\eta_{[1,n]})\right) \nn\\
- J_{F,\kappa_+}\left(\eta_{[1,n]};\cdot,\gamma_{\rtheta_{j},[\kappa_+,n]}^{(\kappa)*}(\eta_{[1,n]})\right)\nn
\end{align}
as the cost in impact since the agent $\rtheta_j$ is in charge of C during $[\kappa,\kappa_+)$ even though he/she has selected his/her strategies as if he/she will be in charge until the end of the horizon.

\noindent
{\bf Remark 5.}
{\em Even though \textup{S} has access to the statistics of the jump process, computation of the expectation \eqref{eq:all0} over all possible sequences $\{\rtheta_1,\ldots,\rtheta_N\}$ is computationally expensive since there are $O(t^N)$ sequences, where $t$ is the number of attackers. Note that most of these sequences have relatively low probability. As an example, multiple successful attacks by different attackers consecutively can be considered as a rare event. Therefore, we let $\Omega \subset \Theta^N$ be the set of selected, typical, sequences. For example, $\Omega$ can include the sequences such that there will be at most one infiltration (by any of the attackers) and the infiltration may or may not be detected until the end of the time horizon. Then, \textup{S} will consider $O(tN^2)$ sequences. Note that each sequence corresponds to a different infiltration scenario. Also let $\mu(\rtheta_{[1,N]})$ denote the normalized measure of a sequence $\rtheta_{[1,N]}\in\Omega$, i.e., $\sum_{\rtheta_{[1,N]}\in\Omega} \mu(\rtheta_{[1,N]}) = 1$.}

Based on Remark 5, the pair \eqref{eq:pair} leads to the Stackelberg equilibrium provided that
\begin{subequations}\label{eq:equilibrium}
\begin{align}
&\hspace{-.1in}\eta^*_{[1,n]} = \argmin_{\substack{\eta_{k}\in\setS,\\ k=1,\ldots,n}}\; \hspace{-.05in}\sum_{\rtheta_{[1,N]}\in\Omega}\hspace{-.1in}\mu(\rtheta_{[1,N]})\hspace{-.2in}\sum_{\kappa \in \hbar(\rtheta_{[1,N]})} \hspace{-.1in}J_{F,\kappa}^{\kappa_+}\left(\eta_{[1,n]};\cdot,\gamma_{\rtheta_{j},[\kappa,n]}^{(\kappa)*}(\eta_{[1,n]})\right) \label{eq:all}\\
&\gamma_{\theta,[\kappa,n]}^{(\kappa)*}(\eta_{[1,n]}) = \argmin_{\substack{\gamma_{\theta,k}^{(\kappa)}\in\setC_k,\\ k=\kappa,\ldots,n}} \; J_{\theta,\kappa}\left(\eta_{[1,n]};\cdot,\gamma_{\theta,[\kappa,n]}^{(\kappa)}(\eta_{[1,n]})\right),
\end{align}
\end{subequations}

\noindent
{\bf Remark 6.}
{\em We note that any brute force approach, trying to solve the optimization problem \eqref{eq:all} numerically (since it is a finite dimensional problem due to linear memoryless \textup{S} strategies), e.g., via particle swarm optimization \cite{PSO}, needs to find $n$ matrices with  $m\times m$ dimensions, corresponding to $nm^2$ parameters, where $m$ is the dimension of the state and $n$ is the number of stages (i.e., time horizon). In particular, we would be searching for a point in $\mathbb{R}^{(nm^2)}$ dimensional space, in addition to the computational load to compute the cost \eqref{eq:all} associated with those points. Furthermore, the result of such a numerical approach would only imply a local optimum, and not the global one.}

In the following sections, we analyze the equilibrium achieving strategies.

\section{Optimal Follower (Controller) Reactions} \label{sec:controller}

For any given S strategies, $\eta_{[1,n]}$, we aim to compute the corresponding reactions $\gamma_{\theta,k}^{(\kappa)}(\eta_{[1,k]})$ for $\theta\in\{\F,\A_1,\ldots,\A_t\}$, $k=\kappa,\ldots,n$, and $\kappa \in \{\delta,\ldots,N\delta\}$. To this end, we first provide friendly C reactions for given sensor strategies and then, we compute adversarial C reactions correspondingly.

\subsection{Optimal Agent-\textup{F} Reaction} \label{sub:F}

Based on Remark 3, secure sensor designer is only interested in the reaction of F when he/she is in charge of C starting from time $k=1$. We note that given linear memoryless S strategies, the problem is an LQG control problem for F \cite{kumarStochasticBook}. In the following, for completeness, we will derive the corresponding optimal control inputs. 

In order to facilitate the subsequent analysis, we can rewrite the state equations \eqref{eq:state}-\eqref{eq:output} and the cost function \eqref{eq:loss} without altering the optimization problem. Particularly, after completing the squares \cite{kumarStochasticBook,Bansal89}, the friendly objective \eqref{eq:loss} is equivalent to:
\begin{equation}\label{eq:loss2}
\min_{\substack{\gamma_{\F,k} \in \setC_k\\k=1,\ldots,n}}\sum_{k=1}^n \E\|\ru_{\F,k} + K_{\F,k} \rx_{\F,k}\|_{\Delta_{\F,k}}^2 + G_{\F},
\end{equation}
where\footnote{Note that $\Delta_{\F,k}$ is invertible since $R_{\F}\succ O$ and $Q_{\F}\succeq O$.}
\begin{subequations}\label{eq:cons2}
\begin{align}
&\Delta_{\F,k} = B' \tQ_{\F,k+1} B + R_{\F},\\
&K_{\F,k} = \Delta_{\F,k}^{-1}B'\tQ_{\F,k+1}A,\\
&G_{\F} = \tr\{\Sigma_{1} (\tQ_{\F,1} - Q_{\F})\} + \sum_{k=1}^n \tr\{\Sigma_v \tQ_{\F,k+1}\}.
\end{align}
\end{subequations}
 The sequence $\{\tQ_{\F,k}\}$ is defined through the following discrete-time Riccati equation:
\begin{subequations}\label{eq:cons22}
\begin{align}
&\tQ_{\F,k} =Q_{\F} + A'\left(\tQ_{\F,k+1} - \tQ_{\F,k+1} B\Delta_{\F,k}^{-1}B'\tQ_{\F,k+1}\right)A,\\
&\tQ_{\F,n+1} = Q_{\F}.
\end{align}
\end{subequations}
Then, through a change of variables \cite{Bansal89}, friendly type C's objective \eqref{eq:loss2} can be written as
\begin{equation}\label{eq:loss3}
\min_{\substack{\gamma_{\F,k} \in \setC_k\\k=1,\ldots,n}}\sum_{k=1}^n \E \|\ru_{\F,k}^{o} + K_{\F,k} \rx_{\F,k}^{o}\|_{\Delta_{\F,k}}^2 + G_{\F}
\end{equation}
subject to \eqref{eq:cons2}-\eqref{eq:cons22} and for $k =1,\cdots,n$,
\begin{subequations}\label{eq:cons3}
\begin{align}
&\rx_{k+1}^{o} =A \rx_k^{o} + \rv_k\mbox{ and }\rx_{1}^{o} = \rx_{1},\label{eq:cons3a}\\
&\ru_{\F,k}^{o} = \ru_{\F,k} + K_{\F,k}B\ru_{\F,k-1} + \cdots + K_{\F,k} A^{k-2}B\ru_{\F,1},\label{eq:cons3b}
\end{align}
\end{subequations}
Note that, now, the process $\{\rx_k^{o}\}$ is independent of how the control inputs $\ru_{\F,k}$ (and $\ru_{\F,k}^{o}$) are constructed while the sensor outputs by depending on the current state $\rx_{\F,k}$ also depend on the previous control inputs. 

Applying the Principle of Optimality to \eqref{eq:loss3}, in view of \eqref{eq:LL}, leads to the result that the last stage optimal transformed control input is given by
\begin{align}
\ru_{\F,n}^{o*} &= -K_{\F,n}\E\{\rx_n^o| \rs_{[1,n]}\}\nn\\
&=-K_{\F,n}\E\{\rx_n^o|\LL_1'\rx_{\F,1},\ldots,\LL_n'\rx_{\F,n}\}.\label{eq:PoO}
\end{align}
However, by \eqref{eq:cons3a}, we have
\begin{equation}\label{eq:lemma1}
\LL_k'\rx_{\F,k} = \LL_k'\rx_k^o + \underbrace{\LL_k'B\ru_{\F,k-1} + \ldots + \LL_k'A^{k-2}B\ru_{\F,1}},
\end{equation}
where the underbraced term is $\sigma$-$\rs_{[1,k-1]}$ measurable, for $k=1,\ldots,n$. Therefore, \eqref{eq:PoO} is equivalent to
\begin{equation}\label{eq:classF}
\ru_{\F,n}^{o*} =-K_{\F,n}\E\{\rx_n^o|\LL_1'\rx_{1}^o,\ldots,\LL_n'\rx_{n}^o\} 
\end{equation}
and does not depend on previous control inputs. Therefore, by induction, we can conclude that the problem entails classical information and for $k=1,\ldots,n$, the optimal transformed control input is given by
\begin{equation}\label{eq:uo}
\ru_{\F,k}^{o*} = - K_{\F,k} \E\{\rx_k^o | \rs_{[1,k]}\}
\end{equation}
almost everywhere on $\mathbb{R}^r$, for $k=1,\ldots,n$, which would also imply the uniqueness of the best F reactions and singleton reaction set $\Pi_{\F,1}$. Then, by \eqref{eq:cons3b} and \eqref{eq:uo}, the optimal control inputs are given by
\begin{align}
\underbrace{\begin{psmallmatrix} \ru_{\F,n}^{*} \\ \vphantom{\int\limits^x}\smash{\vdots} \\ \ru_{\F,1}^{*} \end{psmallmatrix}}_{=:\;\ru_{\F}^*} = - {\underbrace{\begin{psmallmatrix} I_r & K_{\F,n} B & K_{\F,n} A B & \ldots & K_{\F,n} A^{n-2} B \\ & I_r & K_{\F,n-1}B & \cdots & K_{\F,n-1}A^{n-3}B \\ & & I_r & \cdots & K_{\F,n-2}A^{n-4}B \\ & & & \vphantom{\int\limits^x}\smash{\ddots} &  \\ & & & & I_r\end{psmallmatrix}}_{=:\;\Phi_{\F}}}^{-1}\hspace{-.2in}\underbrace{\begin{psmallmatrix} K_{\F,n} & & \\ & \vphantom{\int\limits^x}\smash{\ddots} & \\ & & K_{\F,1} \end{psmallmatrix}}_{=:\;K_{\F}} \hspace{-.05in}\overbrace{\begin{psmallmatrix} \hx_n^o \\  \vphantom{\int\limits^x}\smash{\vdots}\\ \hx_1^o \end{psmallmatrix}}^{=:\;\hx^o},\label{eq:bg}
\end{align}
where $\hx_k^o := \E\{\rx_k^o|\rs_{[1,k]}\}$, and equivalently:
\begin{equation}
\ru_{\F}^* = - \Phi_{\F}^{-1}K_{\F}\hx^o.\label{eq:ustar}
\end{equation}

\subsection{Optimal Adversarial Reaction (Attack)}

Here, we compute the optimal attack strategies with control objective \eqref{eq:lossA}. Different from Subsection \ref{sub:F}, now, we also consider the scenarios where $\A_i$ can infiltrate into C at $k>1$. That would imply that an attacker $\A_j$ can infiltrate into C that has already been infiltrated into by another attacker $\A_i$. Furthermore, the attacker may not know the underlying statistics of the jump process corresponding to the transitions among the agents that can be in charge of C. Correspondingly, the attacker may not know how the state has been driven until he/she has infiltrated into. However, as also noted in Remark 4, consecutive succesful infiltration by different attackers can be considered as a rare event relative to the infiltration of the attacker into C while $\F$ was in charge. Therefore, as an attack model, we consider the scenarios where the attackers can assume that $\F$ was in charge of $\C$ before they have infiltrated into.

Next, we aim to rewrite the state equations and the cost functions as in \eqref{eq:loss2} and \eqref{eq:loss3} for the minimization of the adversarial objectives  \eqref{eq:lossA}. For $k=\kappa,\ldots,n$, let 
\begin{equation}\label{eq:udelta}
\deltau_{\A_i,k} := \ru_{\A_i,k} - \ru_{\F,k}.
\end{equation}
Then, instead of \eqref{eq:state}, consider the following recursion:
\begin{align}
\left[\begin{array}{c}
  \rx_{\A_i,k+1} \\ \hdashline[2pt/2pt]
  \ru_{\F} \\ z_i
\end{array}\right]&
= \underbrace{\left[\begin{array}{c;{2pt/2pt}c}
 A & \begin{array}{cccc} O_{m\times (n-k)r} &B & O_{m\times (k-1)r} & O_m \end{array}\\ \hdashline[2pt/2pt]
  O & I_{m+nr}
\end{array}\right]}_{=:\; \bA_k}\nn\\
\times&\underbrace{\left[\begin{array}{c}
  \rx_{\A_i,k} \\ \hdashline[2pt/2pt]
  \ru_{\F} \\ z_i
\end{array}\right]}_{=\;\bx_{\A_i,k}} + \underbrace{\left[\begin{array}{c}
 \begin{array}{c} B \end{array}\\ \hdashline[2pt/2pt]
  O_{(m+nr)\times r}
\end{array}\right]}_{=: \; \bB} \deltau_{A_i,k} + \underbrace{\left[\begin{array}{c}
  I_m \\ \hdashline[2pt/2pt]
  O
\end{array}\right]}_{=: \; E}\rv_k,\nn
\end{align}
which can be written in compact form as
\begin{equation}\label{eq:stateA}
\bx_{\A_i,k+1} = \bA_k \bx_{\A_i,k} + \bB \deltau_{A_i,k} + E\rv_k.
\end{equation}
Correspondingly, the objective \eqref{eq:lossA} can be rewritten as
\begin{equation}\label{eq:lossA11}
\sum_{k=\kappa}^{n} \E\left\{\|\bx_{\A_i,k+1}\|_{\bQ_{A_i}}^2 + \|\deltau_{A_i,k}\|_{R_{A_i}}^2\right\},
\end{equation}
where 
\begin{equation}
\bQ_{\A_i} := \begin{psmallmatrix} Q_{\A_i} + \lambda_i Q_{\F} & O & -Q_{\A_i}\\
                                                  O & O & O \\
                                                  -Q_{\A_i} & O & Q_{\A_i}\end{psmallmatrix}\succeq O.
\end{equation} 
We point out the resemblance between \eqref{eq:lossA11} and \eqref{eq:loss}. 

As in \eqref{eq:loss2}, we can rewrite the cost function \eqref{eq:lossA11} without altering the optimization problem. After completing the squares, the adversarial objective \eqref{eq:lossA11} is equivalent to:
\begin{equation}\label{eq:lossA2}
\min_{\substack{\gamma_{\A_i,k}^{(\kappa)} \in \setC_k\\k=\kappa,\ldots,n}}\sum_{k=\kappa}^n \Ek\|\deltau_{\A_i,k} + K_{\A_i,k}\bx_{\A_i,k}\|_{\Delta_{\A_i,k}}^2 + G_{\A_i,\kappa},
\end{equation}
where 
\begin{subequations}\label{eq:consA2}
\begin{align}
&\Delta_{\A_i,k} = \bB' \tQ_{\A_i,k+1}\bB + R_{\A_i},\\
&K_{\A_i,k} = \Delta_{\A_i,k}^{-1}\bB'\tQ_{\A_i,k+1}\bA_k,\\
&G_{\A_i,\kappa} = \tr\{\bar{\Sigma}_{\A_i,\kappa} (\tQ_{\A_i,\kappa}-\bQ_{\A_i})\} + \sum_{k=\kappa}^n \tr\{\bar{\Sigma}_v \tQ_{\A_i,k+1}\}, \label{eq:consA2c}
\end{align}
\end{subequations}
and $\bar{\Sigma}_{\A_i,\kappa} := \E\{\bx_{\A_i,\kappa}^{}\bx_{\A_i,\kappa}'\}, \bar{\Sigma}_v := E\Sigma_v E'$. Note that $\bx_{\A_i,\kappa}$ (correspondingly $\bar{\Sigma}_{\A_i,\kappa}$) does not depend on $\A_i$'s strategies $\gamma_{\A_i,[\kappa,n]}^{(\kappa)}$, and instead depends on $\gamma_{\F,[1,\kappa-1]}$. The sequence $\{\tQ_{\A_i,k}\}$ is defined through the following discrete-time Riccati equation:
\begin{align}\label{eq:consA22}
&\tQ_{\A_i,k}=\bQ_{\A_i}+ \bA_k' \Big(\tQ_{\A_i,k+1} -\tQ_{\A_i,k+1}\bB\Delta_{\A_i,k}^{-1} \bB'\tQ_{\A_i,k+1}\Big)\bA_k,\nn\\
&\tQ_{\A_i,n+1} = \bQ_{\A_i}.
\end{align}
We emphasize that $K_{\A_i,k}$, $\Delta_{\A_i,k}$, and $\tQ_{\A_i,k}$ do not depend on the infiltration time $\kappa$.

And corresponding to \eqref{eq:loss3}, the adversarial objective \eqref{eq:lossA2} can be written as
\begin{equation}\label{eq:lossA3}
\min_{\substack{\gamma_{\A_i,k}\in \setC_k\\k=\kappa,\ldots,n}}\sum_{k=\kappa}^n \E \|\deltau_{\A_i,k}^o + K_{\A_i,k}\bx_{\A_i,k}^o\|_{\Delta_{\A_i,k}}^2 + G_{\A_i,\kappa}
\end{equation}
subject to \eqref{eq:consA2}-\eqref{eq:consA22} and for $k=1,\ldots,n$,
\begin{subequations}\label{eq:consA3}
\begin{align}
&\bx_{\A_i,k+1}^o = \bA_k \bx_{\A_i,k}^o + E\rv_k\mbox{ and }\bx_{\A_i,1}^o = \bx_{\A_i,1},\label{eq:consA3a}\\
&\deltau_{\A_i,k}^o = \deltau_{\A_i,k} + K_{\A_i,k}\bB\deltau_{\A_i,k-1} + K_{\A_i,k}\bA_{k-1}\bB\deltau_{\A_i,k-2} + \ldots \nn\\
&+  K_{\A_i,k}\bA_{k-1}\cdots\bA_{\kappa+1}\bB\deltau_{\A_i,\kappa} + K_{\A_i,k}\bA_{k-1}\cdots\bA_{\kappa}\bB\deltau_{\kappa-1} + \ldots\nn\\
&+ K_{\A_i,k} \bA_{k-1}\cdots \bA_2 \bB \deltau_1.\label{eq:consA3b}
\end{align}
\end{subequations}
Note that in \eqref{eq:lossA3}, $G_{\A_i,\kappa}$ does not depend on the adversary's optimization arguments even though it depends on $\ru_{\F}$ due to $\bar{\Sigma}_{\A_i,\kappa}$ in \eqref{eq:consA2c}. However, $\F$ does not consider the impact of $\ru_{\F}$ on $G_{\A_i,\kappa}$ while selecting $\gamma_{\F,[1,n]}$ since \eqref{eq:lossA3} is the cost function of $\A_i$, and not of $\F$. Note also that if $\F$ is in charge of C before $\A_i$ has infiltrated into C, for $k=1,\ldots,\kappa-1$, we have 
\begin{equation}\label{eq:halfzero}
\deltau_{k} = {\bf 0}.
\end{equation}
Even though the process $\{\bx_{\A_i,k}^o\}$ is independent of how the control inputs $\deltau_{\A_i,k}$ (and $\deltau_{\A_i,k}^o$) are constructed, the sensor outputs $\rs_k$, $k=\kappa,\ldots,n$, depend on the taken actions, i.e., $\ru_{\A_i,k}$. 

Similar to \eqref{eq:PoO}, the Principle of Optimality yields
\begin{align}\label{eq:Ailast}
\deltau_{\A_i,n}^{o*} = -K_{\A_i,n}\E\{\bx_{\A_i,k}^o|\LL_1'\rx_{\A_i,1},\ldots,\LL_n'\rx_{\A_i,n}\}.
\end{align}
Then, irrespective of \eqref{eq:consA3a} and \eqref{eq:stateA}, we have 
\begin{align}
\LL_k'\rx_{\A_i,k} = \;&\LL_k'\rx_n^o + \underbrace{\LL_k'B\ru_{\A_i,k-1} + \ldots + \LL_k'A^{k-\kappa-1}B\ru_{\A_i,\kappa}}\nn\\
& \underbrace{+\ldots + \LL_k'A^{k-2}B\ru_{F,1}},\label{eq:lemma2}
\end{align}
where $\rx_k^o$ evolves according to \eqref{eq:cons3a} and the underbraced terms are $\sigma$-$\rs_{[1,k-1]}$ measurable. Therefore, \eqref{eq:Ailast} is equivalent to
\begin{equation}
\deltau_{\A_i,n}^{o*} = -K_{\A_i,n}\E\{\bx_{\A_i,k}^o|\LL_1'\rx_{1}^o,\ldots,\LL_n'\rx_{n}^o\}\label{eq:Aiequ}
\end{equation}
and does not depend on the previous control inputs $\ru_{\A_i,[\kappa,n-1]}$.

\noindent
{\bf Remark 7.} 
{\em We emphasize that if \textup{$\A_i$} does not know which agent was in charge of \textup{C} before $\kappa$, then the relation between $\ru_{[1,\kappa-1]}$ and the corresponding sensor outputs $\rs_{[1,\kappa-1]}$ would not be known by \textup{$\A_i$} and correspondingly \eqref{eq:Aiequ} would not be equivalent of \eqref{eq:Ailast}. Furthermore, the equivalence does also not hold if another attacker had already infiltrated when \textup{$\A_i$} infiltrates since \textup{$\A_i$} has assumed that \textup{$\F$} was in charge before the infiltration. Note also that \textup{$\rmS$} is only interested in the scenarios where $\rtheta_{[1,N]}\in \Omega$, i.e., there is no successful consecutive infiltration by different attackers. A detailed analysis of the consecutive successful infiltration and attacks with partial information is left as future work.}

By induction, we can conclude that the optimal transformed control inputs of $\A_i$ are given by
\begin{equation}\label{eq:deltaustar}
\deltau_{\A_i,k}^{o*} = -K_{\A_i,k} \E\{\bx_{\A_i,k}^o|\rs_{[1,k]}\},
\end{equation}
for $k=\kappa,\ldots,n$, almost everywhere on $\mathbb{R}^{r}$. This implies the uniqueness of the best $\A_i$ reactions and singleton reaction set $\Pi_{\A_i,\kappa}$. Then, by \eqref{eq:consA3b}, we have
\begin{align}
\underbrace{\begin{psmallmatrix}
  \deltau_{\A_i,n}^o \\ \deltau_{\A_i,n-1}^o \\ \vphantom{\int\limits^x}\smash{\vdots} \\ \deltau_{\A_i,\kappa}^o
\end{psmallmatrix}}_{=:\;\deltau_{\A_i}^{(\kappa)o}}& =
\underbrace{\begin{psmallmatrix}
  I_{r} & K_{\A_i,n}\bB & \cdots & K_{\A_i,n}\bA_{n-1}...\bA_{\kappa+1}\bB \\
  & I_{r} & \cdots & K_{\A_i,n-1}\bA_{n-2}...\bA_{\kappa+1}\bB\\
  & & \vphantom{\int\limits^x}\smash{\ddots} & \vphantom{\int\limits^x}\smash{\vdots} \\ & & & I_{r}
\end{psmallmatrix}}_{=: \; \Phi_{\A_i}^{(\kappa)} \in \mathbb{R}^{(n-\kappa+1)r \times (n-\kappa+1)r}}
\underbrace{\begin{psmallmatrix}
  \deltau_{\A_i,n} \\ \deltau_{\A_i,n-1} \\ \vphantom{\int\limits^x}\smash{\vdots} \\ \deltau_{\A_i,\kappa}
\end{psmallmatrix}}_{=: \; \deltau_{\A_i}^{(\kappa)}}\label{eq:phi_Ai}
\end{align}
which can also be written as 
\begin{equation}\label{eq:oneline}
\deltau_{\A_i}^{(\kappa)o} = \Phi_{\A_i}^{(\kappa)} \deltau_{\A_i}^{(\kappa)}
\end{equation} 
while \eqref{eq:deltaustar} leads to
\begin{equation}
\deltau_{\A_i}^{(\kappa)o*} = - \underbrace{\begin{psmallmatrix}
                   K_{\A_i,n} & & \\ & \vphantom{\int\limits^x}\smash{\ddots} & \\ & & K_{\A_i,\kappa}
                 \end{psmallmatrix}}_{=: \; K_{\A_i}^{(\kappa)}} \begin{psmallmatrix}
                   \E\{\bx_{\A_i,n}^o|\rs_{[1,n]}\} \\ \vphantom{\int\limits^x}\smash{\vdots} \\
                   \E\{\bx_{\A_i,\kappa}^o | \rs_{[1,\kappa]}\}
                 \end{psmallmatrix}.\label{eq:KA}
\end{equation}

Next, we seek to compute $\E\{\bx_{\A_i,k}^o|\rs_{[1,k]}\}$ in \eqref{eq:KA} in terms of $\hx^o$. To this end, let us take a closer look at \eqref{eq:consA3a}:
\begin{equation}
\left[\begin{array}{c}
  \rtx_{k+1} \\ \hdashline[2pt/2pt]
  \ru_{\F} \\ z_i
\end{array}\right]
= \left[\begin{array}{c;{2pt/2pt}c}
 A & \begin{array}{cccc} \cdots &B & \cdots & O_m \end{array}\\ \hdashline[2pt/2pt]
  O & I_{m+nr}
\end{array}\right]
\left[\begin{array}{c}
  \rtx_{k} \\ \hdashline[2pt/2pt]
  \ru_{\F} \\ z_i
\end{array}\right] + \left[\begin{array}{c}
  I_m \\ \hdashline[2pt/2pt]
  O
\end{array}\right] \rv_k,\nn
\end{equation}
where we introduce $\rtx_k$, which is given by
\begin{align}
\rtx_k = \rx_k^o + B\ru_{\F,k-1} + AB\ru_{\F,k-2} + \cdots + A^{k-2}B \ru_{\F,1}.\label{eq:xhat2}
\end{align}
Then, we have
\[
\begin{psmallmatrix}
\rtx_n\\ \rtx_{n-1} \\ \vphantom{\int\limits^x}\smash{\vdots} \\ \rtx_1
\end{psmallmatrix} =
\begin{psmallmatrix}
\rx_n^o\\ \rx_{n-1}^o \\ \vphantom{\int\limits^x}\smash{\vdots} \\ \rx_1^o
\end{psmallmatrix} +
\underbrace{\begin{psmallmatrix}
  O & B & AB & \cdots & A^{n-2}B\\
  O & O & B & \cdots & A^{n-3}B \\
  \vphantom{\int\limits^x}\smash{\vdots} & & & & \vphantom{\int\limits^x}\smash{\vdots}\\
  O & O & \cdots & \cdots & O
\end{psmallmatrix}}_{=: \; \Psi}
\begin{psmallmatrix}
\ru_{\F,n}\\ \ru_{\F,n-1} \\ \vphantom{\int\limits^x}\smash{\vdots} \\ \ru_{\F,1}
\end{psmallmatrix}.
\]
Let $\Psi$ be partitioned as $\Psi = [\Psi_n' \cdots \Psi_1']'$ such that $\rtx_k = \rx_k^o + \Psi_k \ru_{\F}$. Therefore, $\E\{\bx_{\A_i,k}^o | \rs_{[1,k]}\}$ can be written as
\begin{equation}
\E\{\bx_{\A_i,k}^o | \rs_{[1,k]}\} = \begin{psmallmatrix} \E\{\rx_k^o | \rs_{[1,k]} \} + \Psi_k \E\{\ru_{\F} | \rs_{[1,k]}\} \\ \E\{\ru_{\F} | \rs_{[1,k]}\} \\ z_i \end{psmallmatrix}.\label{eq:bxo}
\end{equation}

Furthermore, \eqref{eq:ustar} leads to
\begin{equation}
\E\{\ru_{\F} | \rs_{[1,k]}\} = - \Phi_{\F}^{-1}K_{\F}^{} \begin{psmallmatrix}
\E\{\E\{\rx_n^o| \rs_{[1,n]}\} | \rs_{[1,k]}\} \\ \vphantom{\int\limits^x}\smash{\vdots} \\
\E\{\E\{\rx_1^o|\rs_1\} | \rs_{[1,k]}\}
\end{psmallmatrix}.\label{eq:condu}
\end{equation}
Note that we have
\[
\E\{\E\{\rx_l^o | \rs_{[1,l]}\}|\rs_{[1,k]}\} = \left\{\begin{array}{ll}
\E\{\rx_l^o | \rs_{[1,k]}\} & \mbox{if } l\geq k \\
\E\{\rx_l^o | \rs_{[1,l]}\} & \mbox{if } l < k
\end{array}\right.,
\]
where the first case, i.e., $l \geq k$, follows due to the iterated expectations with nested conditioning sets, i.e., $\{\rs_{[1,l]}\} \supseteq \{\rs_{[1,k]}\}$ if $l \geq k$; and the second case, i.e., $l < k$, follows since $\Ek\{\rx_l^o|\rs_{[1,l]}\}$ is $\sigma$-$\rs_{[1,k]}$ measurable if $l < k$. Therefore, \eqref{eq:condu} can be written as
\begin{equation}
\E\{\ru_{\F} | \rs_{[1,k]}\} = -\Phi_{\F}^{-1} K_{\F}^{} \underbrace{\left[\begin{array}{c;{2pt/2pt}c;{2pt/2pt}c}
   O& \begin{array}{c} A^{n-k} \\ \vphantom{\int\limits^x}\smash{\vdots} \\ A \\ I_m\end{array} & O\\ \hdashline[2pt/2pt]
   O & O & I_{(k-1)m}
\end{array}\right]}_{=: \; L_k} \hx^o,
\end{equation}
where the middle block is the $k$th block column from the right. Hence, we can rewrite \eqref{eq:bxo} as
\begin{equation}\label{eq:condx}
\E\{\bx_{\A_i,k}^o | \rs_{[1,k]}\} = \underbrace{\begin{psmallmatrix} E_k - \Psi_k^{} \Phi_{\F}^{-1}K_{\F}^{} L_k^{}  \\ - \Phi_{\F}^{-1}K_{\F}^{} L_k^{} \\ O_m \end{psmallmatrix}}_{=:\; F_k} \hx^o + \underbrace{\begin{psmallmatrix} O_{m\times 1} \\ O_{nr\times 1} \\ z_i \end{psmallmatrix}}_{=:\;\uz_i},
\end{equation}
where $E_k:=\begin{psmallmatrix} O_{m\times (n-k)m} & I_m & O_{m\times (k-1)m} \end{psmallmatrix}$ is the indicator matrix such that $\E\{\rx_k^o|\rs_{[1,k]}\} = E_k \hx^o$, $k=1,\ldots,n$. Then, by \eqref{eq:deltaustar}, \eqref{eq:KA}, and \eqref{eq:condx}, we have
\begin{equation}
\deltau_{\A_i}^{(\kappa)o*} = - K_{\A_i}^{(\kappa)}\left(\underbrace{\begin{psmallmatrix} F_n \\ \vphantom{\int\limits^x}\smash{\vdots} \\ F_{\kappa} \end{psmallmatrix}}_{=: \; F^{(\kappa)}} \hx^o + {\bf 1}_{n-\kappa+1}\otimes \uz_i\right).\label{eq:F}
\end{equation}
Therefore, \eqref{eq:F} and \eqref{eq:oneline} lead to
\begin{align}\nn
\deltau_{\A_i}^{(\kappa)*} = - (\Phi_{\A_i}^{(\kappa)})^{-1}K_{\A_i}^{(\kappa)}\Big(F^{(\kappa)}\hx^o + {\bf 1}_{n-\kappa+1}\otimes\uz_i\Big).
\end{align}
Note that by \eqref{eq:halfzero} and \eqref{eq:ustar}, we obtain
\begin{align}\label{eq:uastar}
\ru_{\A_i}^{(\kappa)*} =\; &- \Big((\Phi_{\A_i}^{(\kappa)})^{-1}K_{\A_i}^{(\kappa)}F^{(\kappa)} + \begin{psmallmatrix}I_{(n-\kappa+1)r} & O_{(n-\kappa+1)r\times (\kappa-1)r} \end{psmallmatrix}\Phi_{\F}^{-1}K_{\F}^{}\Big)\hx^o \nn\\
&- (\Phi_{\A_i}^{(\kappa)})^{-1}K_{\A_i}^{(\kappa)}\big({\bf 1}_{n-\kappa+1}\otimes\uz_i\big).
\end{align}

In the following theorem, we recap these results.

\noindent
{\bf Theorem 1.} {\em Given \textup{S}'s strategies $\eta_{[1,n]}$, \textup{C}'s optimal reactions $\gamma_{\theta,k}^{(\kappa)}(\eta_{[1,k]})$, where \textup{$\theta\in\{\F,\A_1,\ldots,\A_t\}$}, are given by \eqref{eq:ustar} or \eqref{eq:uastar} depending on whether \textup{C} is a friend or an adversary, respectively. These reaction strategies are unique.}

In the following section, we formulate S's optimal strategies.

\section{Optimal Leader (Sensor) Actions}\label{sec:sensor}

For any given $\eta_{[1,n]}$, Theorem 1 provides the unique optimal reactions of friendly and adversarial agents. Now, for each sequence $\theta_{[1,N]} \in \Omega$, we aim to compute the optimal S strategies that minimize
\begin{align}
\sum_{\rtheta_{[1,N]}\in\Omega}\hspace{-.1in}\mu(\rtheta_{[1,N]})\hspace{-.1in}\sum_{\kappa \in \hbar(\rtheta_{[1,N]})} \sum_{k=\kappa}^{\kappa_+ - 1} \Ek\left\{\|\rx_{k+1}\|_{Q_{\F}}^2 + \|\ru_{\rtheta_{j},k}^{(\kappa)*}\|_{R_{\F}}^2\right\}.\label{eq:Ju}
\end{align}
To this end, we first seek to write $\ru_{\theta,k}^{(\kappa)*}$, $\theta\in\{\F,\A_1,\ldots,\A_t\}$, derived in \eqref{eq:ustar} and \eqref{eq:uastar}, in the same form. By \eqref{eq:ustar}, we have
\begin{equation}
\ru_{\F}^* = - \underbrace{\Phi_{\F}^{-1}K_{\F}}_{=:\;T_{\F}^{(1)}}\hx^{o}.\label{eq:us}
\end{equation}
Correspondingly, \eqref{eq:uastar} leads to
\begin{align}
\ru_{\A_i}^{(\kappa)*} &=-\overbrace{(\Phi_{\A_i}^{(\kappa)})^{-1}K_{\A_i}^{(\kappa)}\big({\bf 1}_{n-\kappa+1}\otimes \uz_i \big)}^{=:\;Z_{\A_i}^{(\kappa)}}\nn\\
- &\underbrace{\Big((\Phi_{\A_i}^{(\kappa)})^{-1}K_{\A_i}^{(\kappa)}F^{(\kappa)} + \begin{psmallmatrix}I_{(n-\kappa+1)r} & O_{(n-\kappa+1)r\times (\kappa-1)r} \end{psmallmatrix}\Phi_{\F}^{-1}K_{\F}^{}\Big)}_{=:\;T_{\A_i}^{(\kappa)}}\hx^o.\nn
\end{align}
Let $Z_{\F}^{(1)} = {\bf 0}$ be a zero vector, then for $\theta\in\{\F,\A_1,\ldots,\A_t\}$, we obtain
\begin{equation}
\ru_{\theta}^{(\kappa)} = -T_{\theta}^{(\kappa)}\hx^o - Z_{\theta}^{(\kappa)},\label{eq:u}
\end{equation}
where $T_{\theta}^{(\kappa)}$ is a $(n-\kappa+1)r\times nm$ matrix and $ Z_{\theta}^{(\kappa)} \in \mathbb{R}^{(n-\kappa+1)r}$. However, in \eqref{eq:Ju}, only $\ru_{\rtheta_{j},k}^{(\kappa)*}$ for $k=\kappa,\ldots,\kappa_+-1$ are included. Let $\kappa_T$ be the last non-terminating state transition time, and $n_T$ denote the termination time corresponding to a jump to state $\T$ or end of horizon. Then, by \eqref{eq:u}, for a given realization of the process $\{\rtheta_{j}\}$, e.g., $\theta_{[1,N]}\in\Omega$, we have
\begin{align}
\overbrace{\begin{psmallmatrix} M_{\kappa_T}\ru_{\theta_{j_T}}^{(\kappa_{T})*} \\ \vphantom{\int\limits^x}\smash{\vdots} \\ M_{1}\ru_{\theta_{1}}^{(1)*} \end{psmallmatrix}}^{=:\, \ru_{\theta_{[1,N]}}^*\in\mathbb{R}^{n_Tr}} = -\overbrace{\begin{psmallmatrix}
M_{\kappa_T}T_{\theta_{j_T}}^{(\kappa_{T})} \\ \vphantom{\int\limits^x}\smash{\vdots} \\
M_{1}T_{\theta_1}^{(1)}\end{psmallmatrix}}^{=:\,T_{\theta_{[1,N]}}\in\mathbb{R}^{n_Tr\times nm}}\hx^o - \overbrace{\begin{psmallmatrix} M_{\kappa_T}Z_{\theta_{j_T}}^{(\kappa_T)} \\ \vphantom{\int\limits^x}\smash{\vdots} \\ M_{1}Z_{\theta_{1}}^{(1)} \end{psmallmatrix}}^{=:\,Z_{\theta_{[1,N]}}\in\mathbb{R}^{n_Tr}},\nn
\end{align}
where $M_{\kappa} \in \mathbb{R}^{(\kappa_+-\kappa)r\times (n-\kappa+1)r}$ is given by $ M_{\kappa} = \begin{psmallmatrix} O & I_{(\kappa_+-\kappa)r}\end{psmallmatrix}$. Therefore, we obtain
\begin{equation}
\ru_{\theta_{[1,N]}}^* = - T_{\theta_{[1,N]}}\hx^o - Z_{\theta_{[1,N]}}.\label{eq}
\end{equation}

Even though S constructs a single set of strategies $\{\eta_k\in\setS\}$ without knowing C's type, the resulting sensor outputs $\{\rs_k = \eta_k(\rx_k)\}$ depend on the state $\rx_k$, hence C's type and correspondingly $\theta_{[1,N]}$. However, as shown in Section \ref{sec:controller}, the problem entails classical information and $\hx^o$ does not depend on $\theta_{[1,N]}$. Therefore, let $\ru_{\theta_{[1,N]},k}^*$ be the corresponding control input at time $k$ according to \eqref{eq} for a given realization $\theta_{[1,N]}$. Then, the objective function \eqref{eq:Ju} can be written as
\begin{equation}
\min_{\substack{\eta_k\in\setS,\\ k=1,\ldots,n_T}} \sum_{\rtheta_{[1,N]}\in \Omega} \hspace{-.1in}\mu(\rtheta_{[1,N]}) \sum_{k=1}^{n_T} \E\left\{\|\rx_{\rtheta_{[1,N]},k+1}\|_{Q_{\F}}^2 + \|\ru_{\rtheta_{[1,N]},k}^*\|_{R_{\F}}^2\right\}.\label{eq:Juu}
\end{equation}
Note that the inner summation can be written as
\begin{equation}
\sum_{k=1}^{n_T} \E\|\ru_{\rtheta_{[1,N]},k}^* + K_{\rmS,k}^{[n_T]}\rx_{\rtheta_{[1,N]},k}\|_{\Delta_{\rmS,k}^{[n_T]}}^2 + G_{\rmS}^{[n_T]},\label{eq:inner}
\end{equation}
where
\begin{subequations}
\begin{align}\label{eq:inner_const}
&\Delta_{\rmS,k}^{[n_T]} = B' \tQ_{\rmS,k+1}^{[n_T]}B + R_{\F}\\
&K_{\rmS,k}^{[n_T]} = (\Delta_{\rmS,k}^{[n_T]})^{-1}B' \tQ_{\rmS,k+1}^{[n_T]} A\\
&G_{\rmS}^{[n_T]} = \tr\{\Sigma_1(\tQ_{\rmS,1}^{[n_T]}-Q_{\F})\} + \sum_{k=1}^{n_T}\tr\{\Sigma_v \tQ_{\rmS,k+1}^{[n_T]}\}.
\end{align}
\end{subequations}
Similar to \eqref{eq:cons22}, the sequence $\{\tQ_{\rmS,k}^{[n_T]}\}$ is defined through the following discrete-time Riccati equation:
\begin{align}
&\tQ_{\rmS,k}^{[n_T]} = Q_{\F} + A'(\tQ_{\rmS,k+1}^{[n_T]} - \tQ_{\rmS,k+1}^{[n_T]}B(\Delta_{\rmS,k}^{[n_T]})^{-1}B' \tQ_{\rmS,k+1}^{[n_T]})A,\nn\\
&\tQ_{\rmS,n_T+1} = Q_{\F}.
\end{align}
Then, by \eqref{eq:loss3}, \eqref{eq:inner} leads to
\begin{equation}
\sum_{k=1}^{n_T}\E\|\ru_{\rtheta_{[1,N]},k}^{o*} + K_{\rmS,k}^{[n_T]}\rx_k^o\|_{\Delta_{\rmS,k}^{[n_T]}}^2 + G_{\rmS}^{[n_T]}.\label{eq:newJu}
\end{equation}
subject to \eqref{eq:inner_const}, where $\rx_k^o$ evolves according to \eqref{eq:cons3a}, and
\begin{equation}
\ru_{\rtheta_{[1,N]},k}^{o*} = \ru_{\rtheta_{[1,N]},k}^* + K_{\rmS,k}^{[n_T]}B\ru_{\rtheta_{[1,N]},k-1}^* + \ldots + K_{\rmS,k}^{[n_T]}A^{k-2}B\ru_{\rtheta_{[1,N]},1}^*.\nn
\end{equation}
We point out that due to time-invariant $Q_{\F}$ and $R_{\F}$, we have 
\begin{align}
&\tQ_{\rmS,k}^{[n_T]} = \tQ_{\F,k+n-n_T},\nn\\
&K_{\rmS,k}^{[n_T]} = K_{\F,k+n-n_T},\nn\\
&\Delta_{\rmS,k}^{[n_T]} = \Delta_{\F,k+n-n_T},
\end{align}
where $\tQ_{\F,k},\,K_{\F,k}$, and $\Delta_{\F,k}$ are defined in \eqref{eq:loss3}. Therefore, the summation can be written as
\begin{align}
\sum_{k=1}^{n_T} \|\ru_{\rtheta_{[1,N]},k}^{o*} + &K_{\rmS,k}^{[n_T]} \rx_k^o\|_{\Delta_{\rmS,k}^{[n_T]}}^2 = \E\|\ru_{\rtheta_{[1,N]}}^{o*} + K_{\rmS}^{[n_T]} \rx^o\|_{\Delta_{\rmS}^{[n_T]}}^2,
\end{align}
where 
\begin{align*}
&\Delta_{\rmS}^{[n_T]} := \begin{psmallmatrix} I_{n_Tr} & O_{n_Tr\times (n-n_T)r} \end{psmallmatrix}\begin{psmallmatrix} \Delta_{\F,n} & & \\ &  \vphantom{\int\limits^x}\smash{\ddots} & \\ & & \Delta_{\F,1} \end{psmallmatrix}\begin{psmallmatrix} I_{n_Tr} \\  O_{(n-n_T)r \times n_Tr} \end{psmallmatrix},\\
& K_{\rmS}^{[n_T]} := \begin{psmallmatrix} I_{n_Tr} & O_{n_Tr\times (n-n_T)r} \end{psmallmatrix} K_F \begin{psmallmatrix} I_{n_Tm} \\  O_{(n-n_T)m\times n_Tm} \end{psmallmatrix}\begin{psmallmatrix}  O_{n_Tm\times (n-n_T)m} & I_{n_Tm} \end{psmallmatrix}.
\end{align*}
Furthermore, in terms of $\ru_{\rtheta_{[1,N]}}^{*}$, $\ru_{\rtheta_{[1,N]}}^{o*}$  is given by
\begin{align}
\ru_{\rtheta_{[1,N]}}^{o*} &= \begin{psmallmatrix} I_r & K_{\rmS,n_T}^{[n_T]} B & \ldots & K_{\rmS,n_T}^{[n_T]} A^{n_T-2} B \\ & I_r & \cdots & K_{\rmS,n_T-1}^{[n_T]}A^{n_T-3}B \\  & & \vphantom{\int\limits^x}\smash{\ddots} &  \\  & & & I_r\end{psmallmatrix}\ru_{\rtheta_{[1,N]}}^{*}\nn\\
&=\begin{psmallmatrix} I_{n_Tr} & O_{n_Tr\times (n-n_T)r} \end{psmallmatrix} \Phi_{\F} \begin{psmallmatrix} I_{n_Tr} \\  O_{(n-n_T)r\times n_Tr} \end{psmallmatrix}\ru_{\rtheta_{[1,N]}}^{*}.
\end{align} 

Next, we introduce the parameters:
\begin{align}
\Xi_{\rtheta_{[1,N]}}^{[n_T]} &:= -\begin{psmallmatrix} I & O \end{psmallmatrix} \Phi_{\F} \begin{psmallmatrix} I \\ O \end{psmallmatrix}T_{\rtheta_{[1,N]}}\\
\xi_{\rtheta_{[1,N]}}^{[n_T]} &:= -\begin{psmallmatrix} I & O \end{psmallmatrix} \Phi_{\F} \begin{psmallmatrix} I \\ O \end{psmallmatrix}Z_{\rtheta_{[1,N]}},
\end{align}
almost everywhere on $\mathbb{R}^{rn_T \times mn}$ and $\mathbb{R}^{rn_T}$, respectively, so that
\begin{align}
\ru_{\rtheta_{[1,N]}}^{o*} = \Xi_{\rtheta_{[1,N]}}^{[n_T]} \hx^o + \xi_{\rtheta_{[1,N]}}^{[n_T]}.
\end{align}
Therefore, the objective \eqref{eq:Juu} can be written as
\begin{equation}
\min_{\substack{\eta_k\in\setS,\\ k=1,\ldots,n}} \sum_{\rtheta_{[1,N]}\in \Omega} \hspace{-.1in}\mu(\rtheta_{[1,N]})\left(\E\|\Xi_{\rtheta}\hx^o + \xi_{\rtheta} + K \rx^o\|_{\Delta}^2 + G\right),\label{eq:trobj}
\end{equation}
where for notational simplicity, we let $\Xi_{\rtheta} := \Xi_{\rtheta_{[1,N]}}^{[n_T]}$, $\xi_{\rtheta} := \xi_{\rtheta_{[1,N]}}^{[n_T]}$, $K:=K_{\rmS}^{[n_T]}$, $\Delta := \Delta_{\rmS}^{[n_T]}$, and $G:= G_{\rmS}^{[n_T]}$. The expectation term in \eqref{eq:trobj} yields
\begin{align}
\E\|\Xi_{\rtheta}\hx^o + \xi_{\rtheta} + K \rx^o\|_{\Delta}^2 &= \E\{(\hx^o)'\Xi_{\rtheta}'\Delta \Xi_{\rtheta} \hx^o\} \nn\\
&\hspace{-.6in}+ 2\E\{(\hx^o)'\Xi_{\rtheta}'\Delta K \rx^o\} + \E\{(\rx^o)'K'\Delta K (\rx^o)\}\nn\\
&\hspace{-.6in}+ 2\E\{\xi_{\rtheta}'\Delta(\Xi_{\rtheta}\hx^o + K \rx^o)\} + \E\{\xi_{\rtheta}'\Delta \xi_{\rtheta}\}.\label{eq:long}
\end{align}
Note that $\hx^o$ and $\rx^o$ have zero mean and are independent of $\rtheta_{[1,N]}$ by \eqref{eq:classF} and \eqref{eq:Aiequ}. Correspondingly, $\E\{\xi_{\rtheta}'\Delta(\Xi_{\rtheta}\hx^o + K \rx^o)\} = 0$. Furthermore, the terms 
\[
\E\{(\rx^o)'K'\Delta K (\rx^o)\} = \tr\{\E\{\rx^o(\rx^o)'\}K'\Delta K\}
\] 
and 
\[
\E\{\xi_{\rtheta}'\Delta \xi_{\rtheta}\} = \tr\{\xi_{\rtheta}\xi_{\rtheta}'\Delta\}
\]
do not depend on the optimization arguments $\eta_{[1,n]}$. Therefore, \eqref{eq:long} yields
\begin{align}
\E\|\Xi_{\rtheta}\hx^o + \xi_{\rtheta} + K \rx^o\|_{\Delta}^2 &= \tr\left\{\E\{\hx^o(\hx^o)'\}\Xi_{\rtheta}'\Delta\Xi_{\rtheta}\right\}\nn\\
&\hspace{-0.6in}+\tr\{\E\{\rx^o(\hx^o)'\}\Xi_{\rtheta}'\Delta K\} + \tr\{\E\{\hx^o(\rx^o)'\}K'\Delta\Xi_{\rtheta}\}\nn\\
&\hspace{-0.6in}+ \tr\left\{\Ek\{\rx^o(\rx^o)'\}K'\Delta K\}+\tr\{\xi_{\rtheta}\xi_{\rtheta}'\Delta\right\}.\label{eq:longlong}
\end{align}

We aim to compute $\E\{\hx^o(\hx^o)'\}$, $\E\{\rx^o(\hx^o)'\}$, and $\E\{\rx^o(\rx^o)'\}$. Let $\Sigma_k^o := \E\{\rx_k^o(\rx_k^o)'\}$. Then, we have
\begin{equation}
\Sigma^o := \E\{\rx^o(\rx^o)'\} =
\begin{psmallmatrix}
\Sigma_n^o & A \Sigma_{n-1}^o & \cdots & A^{n-1}\Sigma_1^o \\
\Sigma_{n-1}^oA' & \Sigma_{n-1}^o & \cdots & A^{n-2}\Sigma_1^o\\
\vphantom{\int\limits^x}\smash{\vdots} & \vphantom{\int\limits^x}\smash{\vdots} & \vphantom{\int\limits^x}\smash{\ddots} & \vphantom{\int\limits^x}\smash{\vdots} \\
\Sigma_1^o (A^{n-1})' & \Sigma_1^o(A^{n-2})' & \cdots & \Sigma_1^o
\end{psmallmatrix}.
\end{equation}
Furthermore, let $H_k := \E\{\hx_k^o(\hx_k^o)'\}$. Then, $\E\{\hx^o(\hx^o)'\}$ can be written as
\begin{align}
\E\{\hx^o(\hx^o)'\} &=
\begin{psmallmatrix}
\E\{\hx_n^o(\hx_n^o)'\} & \cdots & \E\{\hx_n^o(\hx_1^o)'\}\\
\vphantom{\int\limits^x}\smash{\vdots} &
\vphantom{\int\limits^x}\smash{\ddots} &
\vphantom{\int\limits^x}\smash{\vdots}\\
\E\{\hx_1^o(\hx_n^o)'\} & \cdots & \E\{\hx_1^o(\hx_1^o)'\}
\end{psmallmatrix}\nn\\
&= \begin{psmallmatrix}
H_n & AH_{n-1} & \cdots & A^{n-1}H_1 \\
H_{n-1}A'& H_{n-1} & \cdots & A^{n-2}H_1\\
\vphantom{\int\limits^x}\smash{\vdots} &
\vphantom{\int\limits^x}\smash{\vdots} &
\vphantom{\int\limits^x}\smash{\ddots} &
\vphantom{\int\limits^x}\smash{\vdots}\\
H_1(A^{n-1})' & H_1(A^{n-2})' & \cdots & H_1
\end{psmallmatrix}
\end{align}
since for $l < k$, we have
\begin{align}
\E\{\hx_l^o(\hx_k^o)'\} &= \E\{\E\{\hx_l^o(\hx_k^o)'|\rs_{[1,l]}\}\}\nn\\
&\stackrel{(a)}{=} \E\{\hx_l^o \E\{\hx_k^o|\rs_{[1,l]}\}\}\nn\\
&\stackrel{(b)}{=} \E\{\hx_l^o(\hx_l^o)'\}(A^{k-l})',\nn
\end{align}
where $(a)$ holds since $\hx_l^o$ is $\sigma$-$\rs_{[1,l]}$ measurable, and $(b)$ follows due to the iterated expectations with nested conditioning sets, i.e., $\{\rs_{[1,l]}\}\subseteq\{\rs_{[1,k]}\}$. We also note that for $l \leq k$, $\E\{\hx_l^o(\rx_k^o)'\} = \E\{\hx_l^o(\rx_l^o)'\}(A^{k-l})'$ since $\rv_j$, $j>l$, and $\hx_l^o$, which is $\sigma$-$\rs_{[1,l]}$ measurable, are independent of each other and $\{\rv_k\}$ is a zero-mean white noise process. This leads to 
\begin{align}
\E\{\hx_l^o(\rx_l^o)'\} &= \E\{\E\{\hx_l^o(\rx_l^o)'| \rs_{[1,l]}\}\}\nn\\
&= \E\{\hx_l^o(\hx_l^o)'\}\nn
\end{align}
due to the law of iterated expectations. This implies that 
\begin{align}
\E\{\hx_l^o(\rx_k^o)'\} &= \E\{\hx_l^o(\hx_l^o)'\}(A^{k-l})' \nn\\
&= \E\{\hx_l^o(\hx_k^o)'\}\nn
\end{align} 
and correspondingly $\E\{\hx^o(\rx^o)'\} = \E\{\hx^o(\hx^o)'\}$.

Next, we can rewrite \eqref{eq:trobj} as
\begin{equation}
\min_{\substack{\eta_k\in\setS,\\ k=1,\ldots,n}} \tr\left\{\begin{psmallmatrix}
H_n & AH_{n-1} & \cdots & A^{n-1}H_1 \\
H_{n-1}A'& H_{n-1} & \cdots & A^{n-2}H_1\\
\vphantom{\int\limits^x}\smash{\vdots} &
\vphantom{\int\limits^x}\smash{\vdots} &
\vphantom{\int\limits^x}\smash{\ddots} &
\vphantom{\int\limits^x}\smash{\vdots}\\
H_1(A^{n-1})' & H_1(A^{n-2})' & \cdots & H_1
\end{psmallmatrix}\Pi\right\} + \Pi_o, \label{eq:vk}
\end{equation}
where
\begin{subequations}\label{eq:Pi}
\begin{align}
&\Pi := \hspace{-.1in}\sum_{\rtheta_{[1,N]}\in \Omega} \hspace{-.1in}\mu(\rtheta_{[1,N]})\{\Xi_{\rtheta}'\Delta\Xi_{\rtheta} + \Xi_{\rtheta}'\Delta K + K'\Delta \Xi_{\rtheta}\},\\
&\Pi_o := \hspace{-.1in}\sum_{\rtheta_{[1,N]}\in \Omega} \hspace{-.1in}\mu(\rtheta_{[1,N]})(\tr\{\Sigma^o K'\Delta K\} + \tr\{\xi_{\rtheta}\xi_{\rtheta}'\Delta\} + G),
\end{align}
\end{subequations}
which are independent of the optimization arguments. Hence, the optimization problem \eqref{eq:vk} faced by S can be written as an affine function of $H_k$'s as follows:
\begin{equation}
\min_{\substack{\eta_k\in\setS,\\ k=1,\ldots,n}} \sum_{k=1}^n \tr\{V_k H_k\} + \Pi_o,\label{eq:sdp}
\end{equation}
for certain symmetric deterministic matrices\footnote{$\mathbb{S}^m$ denotes the set of $m\times m$ symmetric matrices.} $V_k \in \mathbb{S}^{m}$, $k=1,\ldots,n$, which are given by
\begin{equation}
V_k := \Pi_{k,k} + \sum_{l=k+1}^n \Pi_{k,l}A^{l-k} + (A^{l-k})'\Pi_{l,k},\label{eq:v}
\end{equation}
and $\Pi_{k,l}$ is the corresponding $m\times m$ sub-block of $\Pi$.

As a secure sensor designer, we seek to solve this nonlinear optimization problem \eqref{eq:sdp}. Note that $H_k = \E\{\hx_k^o(\hx_k^o)'\}$ and $\hx_k^o = \E\{\rx_k^o|\rs_{[1,k]}\}$. However, as pointed out in Remark 6, a brute force approach is computationally expensive. To this end, we employ the approach in \cite{sayinAutomatica17}, which considers another optimization problem that bounds the original one from below, and then, compute strategies for the original problem, which optimize the lower bound. Based on this, the following theorem characterizes equilibrium achieving secure sensor strategies.

\begin{table}[t!]
\renewcommand{\arraystretch}{1.4}
  \caption{Computation of secure sensor strategies.}
  \centering
  \begin{tabularx}{.47\textwidth}{ l}
          \hline
          \textbf{Algorithm 1:} Secure Sensor Design \\
          \hline
          \textbf{Compute $V_k$'s:}\\
          \hspace{1em} {\em Compute \textup{$K_{\F,k},\Delta_{\F,k}$}, and \textup{$K_{\A_i,k}$} for $k=1,\ldots,n$ and  $i=1,\ldots,t$}\\
          \hspace{2em} {\em via \eqref{eq:cons2} and \eqref{eq:consA2}.}\\
          \hspace{1em} {\em Compute \textup{$\Phi_{\F}$} by \eqref{eq:bg}, \textup{$\Phi_{\A_i}^{(1)}$} by \eqref{eq:phi_Ai}, and $F^{(1)}$ by \eqref{eq:F}.}\\
          \hspace{1em} {\em Compute $\Pi$ and $\Pi_o$, given by \eqref{eq:Pi}, by computing $\Xi_{\theta}$ and $\xi_{\theta}$}\\
          \hspace{2em} {\em for all $\rtheta_{[1,N]}\in \Omega$.}\\
          \hspace{1em} {\em Then, compute $V_k$, $k=1,\ldots,n$, via \eqref{eq:v}.}\\
          \textbf{SDP Problem:} \\
          \hspace{1em} {\em Solve the SDP problem on the left hand side of \eqref{eq:lowerbound} through}\\
          \hspace{2em} {\em a numerical toolbox, e.g., CVX \cite{cvx,gb08}, and obtain the}\\
          \hspace{2em} {\em solutions $S_k^*$, for $k=1,\ldots,n$.}\\
          \hspace{1em} {\em Set $S_{0}^* = O$.}\\
          \textbf{Equilibrium achieving sensor strategies:}\\
          \hspace{1em} {\em Compute the corresponding idempotent matrices $P_k$,$\forall k$,  by}\\
          \hspace{2em} {\em using $S_k^*$, $\forall k$, and \eqref{eq:solution}.}\\
          \hspace{1em} {\em Compute the eigen decompositions: $P_k = U_k\Lambda_kU_k'$.}\\
          \hspace{1em} {\em Compute $\LL_k$, $\forall k$, by using $S_{k-1}^*,U_k,\Lambda_k$, and \eqref{eq:ck}.}\\
          \hspace{1em} {\em And $\eta_k(\rx_k) = \LL_k'\rx_k$.}\\
          \hline
  \end{tabularx}\label{table}
\end{table}

\noindent
{\bf Theorem 2.} {\em The optimal linear secure sensor strategies can be computed via Algorithm 1, described in Table \ref{table}.}

{\em Proof.}
Based on Lemma 3 in \cite{sayinAutomatica17}, by characterizing necessary conditions on $H_k$'s, we have
\begin{equation}\label{eq:lowerbound}
\begin{array}{ccc}  \min_{\substack{S_k\in\mathbb{S}^m,\\ k=1,\ldots,n}} \sum_{k=1}^n \tr\{V_k S_k\}
& \leq & \min_{\substack{\eta_k\in\setS,\\ k=1,\ldots,n}} \sum_{k=1}^n \tr\{V_k H_k\},\\
\mathrm{s.t. }\;\;\; \Sigma_j^o \succeq S_j \succeq AS_{j-1}A' \; \forall j & & \end{array}
\end{equation}
where $\Sigma_j^o = \Ek\{\rx_j^o(\rx_j^o)'\}$, $S_0 := O$. Note that the left hand side of \eqref{eq:lowerbound} is an SDP problem. By invoking Theorem 4 in \cite{sayinAutomatica17}, we can characterize the solution, $S_1^*,\ldots,S_n^*$, as
\begin{equation}\label{eq:solution}
S_k^* = AS_{k-1}^*A' + (\Sigma_k^o -AS_{k-1}^* A')^{1/2}P_k (\Sigma_k^o - AS_{k-1}^*A')^{1/2},
\end{equation}
for $k=1,\ldots,n$, where $S_0^* = O$ and $P_k \in \mathbb{S}^{m}$ is a certain symmetric idempotent matrix. Note that by solving the SDP problem numerically, we can compute the corresponding $P_k$'s.

Next, say that S employs memoryless linear policies $\rs_k = \eta_k(\rx_{k}) = \LL_k' \rx_{k}$. Then, by \eqref{eq:lemma1} and \eqref{eq:lemma2}, we have
\begin{equation}\nn
\hx_k^o = \E \{\rx_k^o | \LL_1'\rx_1^o,\ldots,\LL_k'\rx_k^o\}.
\end{equation}
which can also be written as
\begin{align}\nn
\hx_k^o = A\hx_{k-1}^o + (&\Sigma_k^o - AH_{k-1}A')\LL_k(\LL_k'(\Sigma_k^o - AH_{k-1}A')\LL_k)^{\dag}\nn\\
&\times \LL_k'(\rx_k^o - A\hx_{k-1}^o),
\end{align}
for $k=1,\ldots,n$, $\hx_{-1}^o := 0$ and $H_0 := O$. Therefore, $H_k = \E\{\hx_k^o(\hx_k^o)'\}$ is given by
\begin{align}\label{eq:solution2}
H_k = AH_{k-1}A' + (&\Sigma_k^o - AH_{k-1}A')\LL_k(\LL_k'(\Sigma_k^o - AH_{k-1}A')\LL_k)^{\dag}\nn\\
&\times \LL_k'(\Sigma_k^o - AH_{k-1}A').
\end{align}
We emphasize the resemblance between \eqref{eq:solution} and \eqref{eq:solution2}.
In particular, if we set $\bar{\LL}_k := (\Sigma_k^o - AH_{k-1}A')^{1/2}\LL_k$, $k=1,\ldots,n$, \eqref{eq:solution2} yields
\begin{align}
H_k = AH_{k-1}A' + (&\Sigma_k^o - AH_{k-1}A')^{1/2}\bar{\LL}_k(\bar{\LL}_k'\bar{\LL}_k)^{\dag}\bar{\LL}_k'\nn\\
&\times(\Sigma_k^o - AH_{k-1}A')^{1/2},
\end{align}
where $\bar{\LL}_k(\bar{\LL}_k'\bar{\LL}_k)^{\dag}\bar{\LL}_k'$ is also a symmetric idempotent matrix just like $P_k$ in \eqref{eq:solution}.

Therefore, given $P_k$'s, let $P_k = U_k\Lambda_kU_k'$ be the eigen decomposition, and set $\bar{\LL}_k = U_k\Lambda_k$, i.e., set
\begin{equation}\label{eq:ck}
\LL_k = (\Sigma_k^o - AS_{k-1}^*A')^{-1/2}U_k\Lambda_k.
\end{equation}
Then, we obtain $H_k = S_k^*$, which implies that S's optimal strategies are given by \eqref{eq:ck} while the optimal control inputs for both friendly and adversarial C are given by \eqref{eq:ustar} or \eqref{eq:uastar}, respectively. \hfill $\square$

\noindent
{\bf Remark 8.}
{\em We have considered that \textup{S} knows the underlying state transition probabilities, i.e., the statistics of the jump process $\{\rtheta_{j}\}$. However, for a robust design against inaccurate perception of the statistics, \textup{S} can design the sensor outputs by being cautious for the worst case scenario. For all possible measures of $\rtheta_{[1,N]}\in\Omega$, we can recompute $\Pi$ and $\Pi_o$ in \eqref{eq:Pi}. Let $\vv$ denote the set of the corresponding $(V_1,\ldots,V_n)$ tuples, which can be computed via  \eqref{eq:v}. Then, by \eqref{eq:lowerbound}, the worst case scenario is equivalent to
\begin{equation}
\max_{V_{[1,n]}\in\vv} \min_{\substack{S_k\in\mathbb{S}^m,\\ k=1,\ldots,n}} \sum_{k=1}^n \tr\{V_k S_k\}
\end{equation}
subject to $\Sigma_j \succeq S_j \succeq AS_{j-1}A'$, $j=1,\ldots,n$ and $S_0 = O$. If $\vv$ is compact and convex, there exists a saddle point equilibrium by the Minimax theorem \cite{basarDynamicBook}. Otherwise, we can consider the convex hull of $\vv$, denoted by $\vv^c$ for computational simplicity in addition to robustness. A detailed analysis in that direction is left as future work.}

In the following section, we provide several numerical examples examining the performance of secure sensor design.

\section{Illustrative Examples} \label{sec:examples}

As numerical illustrations, we consider two different scenarios: Scenario 1, where we compare the performance of the proposed secure sensors with classical sensors that disclose $\rx_k$ to C directly, and Scenario 2, where we analyze the robustness of the proposed scheme against inaccurate perception of the state transition statistics, i.e., $\mu_P(\theta_{[1,N]})\neq\mu(\theta_{[1,N]})$ for $\theta_{[1,N]}\in\Omega$, where $\mu_P$ denotes the perceived statistics while $\mu$ denotes the actual ones. We set time horizon $n = 100$, the state's dimension $m=8$, and the control input's dimension $r = 2$. The matrices in the state recursion \eqref{eq:state} are set randomly according to uniform distribution such that $A$ is not a singular matrix and scaled by $0.1$ in order to avoid computational instability. In order to construct auto-covariance matrices $\Sigma_1$ and $\Sigma_v$, we draw a number from the uniform distribution on the unit interval $[0, 1]$ for each entry of a matrix $D$. Then, we can construct a positive-definite covariance matrix by $(D+D')/2 + 2m\,I$, where the last term ensures that the constructed matrix is diagonally dominant, and therefore, positive-definite. Then, we have scaled $\Sigma_v$ by $10$ since the sensor outputs play more essential role for C when the state noise variance is larger. 

\begin{figure}[t!]
  \centering
  \includegraphics[width=.3\textwidth]{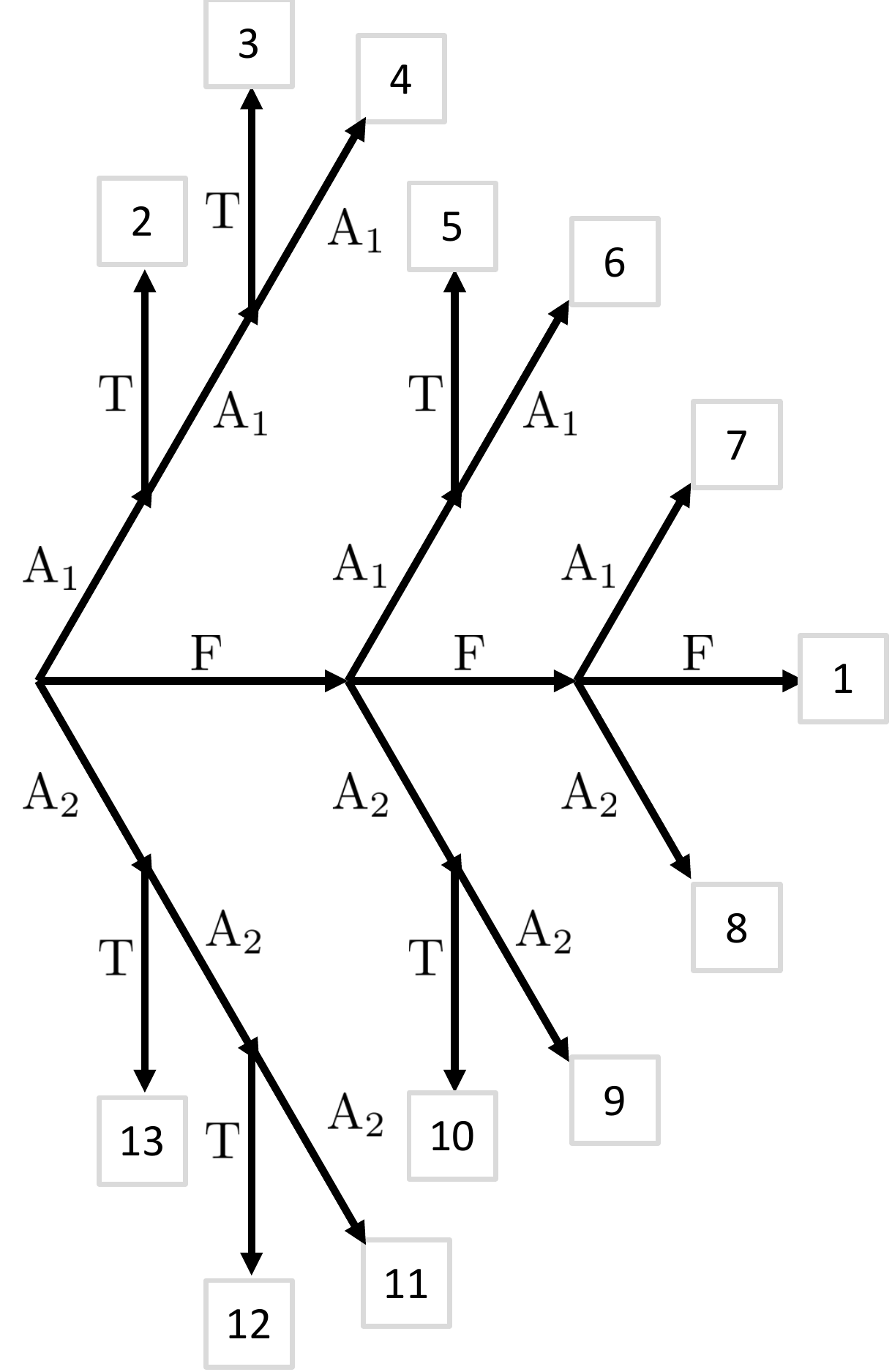}\\
  \caption{Possible state transitions within the time horizon.}\label{fig:two}
\end{figure}

{\em Scenario 1 - Performance Comparison.}
We specifically consider the scenario where there are two adversaries, who seek to regularize state $\rx_k$ around $z_1,z_2 \in \mathbb{R}^m$ that are drawn from a multivariate Gaussian distribution. The positive semi-definite weight matrices in the cost functions are set such that 
\begin{equation}
Q_{\F} = \begin{psmallmatrix} I_{m/2} & O \\ O & O\end{psmallmatrix}, 
Q_{\A_1} = \begin{psmallmatrix} O & O \\ O & I_{m/4}\end{psmallmatrix}, 
Q_{\A_2} = \begin{psmallmatrix} O & O \\ O & I_{m/2}\end{psmallmatrix}, 
\end{equation} 
and $R_{\F} = R_{\A_1} = R_{\A_2} = I_r$ while $\lambda_1 = \lambda_2 = 0.1$. The state transition interval is set $\delta = 35$ and Fig. \ref{fig:two} shows the possible state transitions, e.g., $\rtheta_{[1,N]}\in\Omega$, within the time horizon. We consider the situation where the normalized measure of no-infiltration case, i.e., $\theta_{[1,N]} = \{\F,\F,\F\}$, referred to as Case-$1$, is 0.7 while all the other cases, i.e., Cases $2$-$13$, have the same measure, as tabulated in Table \ref{tab:two}.

\begin{table}[t!]
\renewcommand{\arraystretch}{1.2}
  \caption{Comparison of $\hat{J}_{\rmS}$, defined in \eqref{eq:hatJ}, of different schemes for the cases shown in Fig. \ref{fig:two}.}
  \centering
  \begin{tabular}{ccccc}
     \hline
     \multirow{ 2}{*}{Cases} & \multirow{ 2}{*}{Probability} & \multirow{ 2}{*}{Classical} & \multirow{ 2}{*}{Secure} & No Sensor\\
     & & & & Output \\
     \hline\hline
     $1$ & $0.700$ & $\mathbf{0}$ & $0.1$ & $476$ \\\hline
     $2$ & $0.025$ & $10.5$ & $\mathbf{1.9}$ & $295.3$ \\\hline
     $3$ & $0.025$ & $30.7$  & $\mathbf{13.2}$ & $609.0$ \\\hline
     $4$ & $0.025$ & $57.2$ & $\mathbf{32.0}$ & $886.9$ \\\hline
     $5$ & $0.025$ & $11.3$  & $\mathbf{2.5}$ & $604.6$ \\\hline
     $6$ & $0.025$ & $37.8$ & $\mathbf{21.3}$ & $882.4$ \\\hline
     $7$ & $0.025$ & $17.6$ & $\mathbf{10.0}$ & $877.8$ \\\hline
     $8$ & $0.025$ & $32.2$ & $\mathbf{20.7}$ & $884.7$ \\\hline
     $9$ & $0.025$ & $68.9$ & $\mathbf{43.8}$ & $896.9$ \\\hline
     $10$ & $0.025$ & $27.8$  & $\mathbf{14.3}$ & $612.2$ \\\hline
     $11$ & $0.025$ & $104.3$ & $\mathbf{66.2}$ & $908.7$ \\\hline
     $12$ & $0.025$ & $63.2$ & $\mathbf{36.7}$ & $624.0$ \\\hline
     $13$ & $0.025$ & $26.5$ & $\mathbf{13.6}$ & $302.6$ \\
     \hline\hline
      & Average & $12.2$ & $\mathbf{7.0}$ & $821.2$\\
     \cline{2-5}
   \end{tabular}\label{tab:two}
\end{table}

In Table \ref{tab:two}, we compare the performances of three different schemes for the cases shown in Fig. \ref{fig:two}. The classical scheme refers to a sensor, who discloses $\rx_k$ directly to C, while no sensor output refers to open-loop control of the system, i.e., $\rs_k = {\bf 0}$ almost everywhere on $\mathbb{R}^m$. As a performance measure for each $\rtheta_{[1,N]}\in\Omega$, we consider \eqref{eq:inner}, where the last term $G_{\rmS}^{[n_T]}$, which does not depend on S's strategies $\eta_{[1,n]}$, is excluded, i.e.,
\begin{equation}\label{eq:hatJ}
\hat{J}_{\rmS}(\rtheta_{[1,N]}) := \sum_{k=1}^{n_T} \E\|\ru_{\rtheta_{[1,N]}} + K_{\rmS,k}^{[n_T]}\rx_{\rtheta_{[1,N]},k}\|_{\Delta_{\rmS,k}^{[n_T]}}^2
\end{equation}
and, on the average,
\begin{equation}\label{eq:Jave}
\sum_{\rtheta_{[1,N]}\in\Omega}\mu(\rtheta_{[1,N]}) \hat{J}_{\rmS}(\rtheta_{[1,N]}).
\end{equation}
In Table \ref{tab:two}, we have observed that open-loop control of the system leads to inferior performance compared to the other schemes in all the cases. The classical scheme outperforms the proposed secure sensor scheme only in Case $1$ yet slightly. Note that Case $1$ is the best case scenario, where there is no infiltration into C. In all the other cases, the proposed scheme outperforms the classical scheme substantially. Even though the best case scenario is relatively likely compared to the other cases, on the average, the proposed scheme outperforms the classical scheme also substantially and achieves $40\%$ enhancement in the performance.

\begin{figure}[t!]
  \centering
  \includegraphics[width=.45\textwidth]{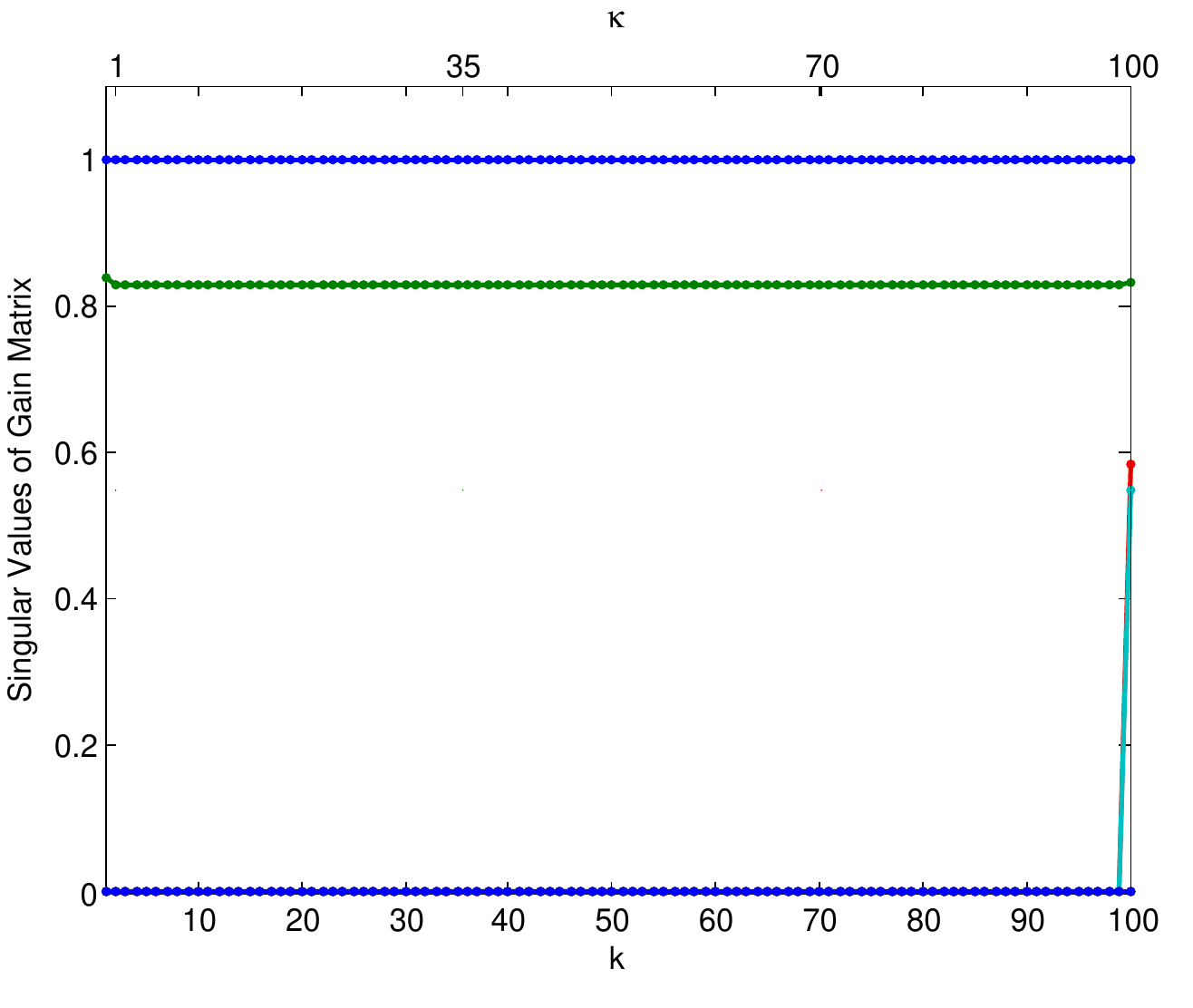}\\
  \caption{Singular values of the normalized gain matrix $\frac{\LL_k}{\|\LL_k\|}$. $\LL_n$ has rank $4$ while all the others $\LL_1,\ldots,\LL_{n-1}$ have rank $2$.}\label{fig:svd}
\end{figure}

Furthermore, in Fig. \ref{fig:svd}, we have plotted the time evolution of the singular values of the normalized gain matrix $\frac{\LL_k}{\|\LL_k\|}$. Note that the gain matrix $\LL_k$ and the normalized gain matrix $\frac{\LL_k}{\|\LL_k\|}$ leads to the same performance, while the normalized one is preferred in Fig. \ref{fig:svd} for better demonstration. We also note that while $\LL_k \in \mathbb{R}^{m\times m}$, all the gain matrices have rank less than $m=8$, i.e., $\LL_1,\ldots,\LL_{n-1}$ have rank $2$ and the last stage gain matrix $\LL_n$ has rank $4$. 

\begin{table}[t!]
\renewcommand{\arraystretch}{1.2}
  \caption{Comparison of $\hat{J}_{\rmS}$ in terms of $\mu$ and $\mu_P$.}
  \centering
  \begin{tabular}{c|cc|cc}
     \hline
     \multirow{ 2}{*}{Cases} & Actual & Accurate & Perceived & Inaccurate\\
     & Probabilities & Perception &  Probabilities & Perception \\
     \hline\hline
     $1$ & $0.700$ & $0.1$       & $0.85$ & $0.0$ \\\hline
     $2$ & $0.025$ & $1.9$ & $0.025$ & $1.9$ \\\hline
     $3$ & $0.025$ & $13.2$ & $0.025$ & $13.3$ \\\hline
     $4$ & $0.025$ & $32.0$ & $0.025$ & $32.1$ \\\hline
     $5$ & $0.025$ & $2.5$ & $0.025$ & $2.5$ \\\hline
     $6$ & $0.025$ & $21.3$ & $0.025$ & $21.3$ \\\hline
     $7$ & $0.025$ & $10.0$ & $0.025$ & $10.0$ \\\hline
     $8$ & $0.025$ & $20.7$ & $0$ & $24.8$ \\\hline
     $9$ & $0.025$ & $43.8$ & $0$ & $53.0$ \\\hline
     $10$ & $0.025$ & $14.3$ & $0$ & $19.3$ \\\hline
     $11$ & $0.025$ & $66.2$ & $0$ & $80.0$ \\\hline
     $12$ & $0.025$ & $36.7$ & $0$ & $46.3$ \\\hline
     $13$ & $0.025$ & $13.6$ & $0$ & $18.2$ \\
     \hline\hline
      & Average & $7.0$ & Average & $8.1$\\
     \cline{2-5}
   \end{tabular}\label{tab:robustness}
\end{table}

{\em Scenario 2 - Robustness Analysis.}
We examine the robustness of the proposed scheme for inaccurate perception of the underlying state transition statistics. To this end, we consider the setup in Scenario 1; however, now S perceives the statistics as tabulated in Table \ref{tab:robustness}. Particularly, S is not aware of the attacker $\A_2$, and designs the secure sensor outputs accordingly even though the actual underlying statistics are as in Scenario 1, which is also provided in Table \ref{tab:robustness}. We have observed that the performance degrades due to inaccurate perception of the statistics; however, the proposed scheme still outperforms classical scheme, whose performance is tabulated in Table \ref{tab:two}, in individual cases, Cases $2-13$, and on the average. We note that since, in the inaccurate perception of the statistics, Case $1$ has relatively higher probability, i.e., $0.85>0.70$, we have observed slight improvement in the performance in that case.

\section{Conclusion} \label{sec:conclusion}

In this paper, we have proposed and addressed secure sensor design problem for cyber-physical systems with linear Gaussian dynamics against the advanced persistent threats with quadratic control objectives. By designing sensor outputs cautiously in advance, we have sought to minimize the damage that can be caused by undetected target-specific threats. To this end, we have modeled the problem formally in a game-theoretical setting. We have determined the optimal control inputs for both friendly and adversarial objectives for given linear sensor strategies. Then, we have provided an algorithm to compute the optimal linear secure sensor strategies that lead to the equilibrium. We note that without linearity assumption on the sensor outputs, the problem entails non-classical information model due to distinct objectives of the agents. Furthermore, for general, e.g., nonlinear, sensor outputs, the corresponding optimal control policies could not be unique and could not even be expressed in closed form \cite{kumarStochasticBook}.

Some future directions of research on this topic include: $\bullet$ Formulation of secure sensor design strategies when the sensor has access to noisy observations, or for, e.g., robust control or feedback stability of the systems. $\bullet$ Here, we have considered the scenarios, where the attackers have perfect knowledge about the underlying state recursion. Another interesting extension would be to analyze the scenarios, where the attackers can only have partial knowledge. In such scenarios, intuitively, sensor outputs could play relatively more powerful roles since, without caution, sensors might help the attackers to recover the unknown part of the system dynamics.

\bibliographystyle{IEEEtran}
\bibliography{sayin}

\begin{IEEEbiographynophoto}{Muhammed O. Sayin}  is currently pursuing the Ph.D. degree in Electrical and Computer Engineering from the University of Illinois at Urbana-Champaign (UIUC). He received the B.S. and M.S. degrees in Electrical and Electronics Engineering from Bilkent University, Ankara, Turkey, in 2013 and 2015, respectively. His current research interests include signaling games, dynamic games and decision theory, and cyber-physical systems.
\end{IEEEbiographynophoto}

\begin{IEEEbiographynophoto}{Tamer Ba\c{s}ar}
is with the University of Illinois at Urbana-Champaign, where he holds the academic positions of  Swanlund Endowed Chair; Center for Advanced Study Professor of  Electrical and Computer Engineering; Research Professor at the Coordinated Science Laboratory; and Research Professor  at the Information Trust Institute. He is also the Director of the Center for Advanced Study.

He received B.S.E.E. from Robert College, Istanbul,
and M.S., M.Phil, and Ph.D. from Yale University. He is a member of the US National Academy
of Engineering,  member of the  European Academy of Sciences, and Fellow of IEEE, IFAC (International Federation of Automatic Control) and SIAM (Society for Industrial and Applied Mathematics), and has served as president of IEEE CSS (Control Systems  Society), ISDG (International Society of Dynamic Games), and AACC (American Automatic Control Council). He has received several awards and recognitions over the years, including the highest awards of IEEE CSS, IFAC, AACC, and ISDG, the IEEE Control Systems Award, and a number of international honorary doctorates and professorships. He has over 800 publications in systems, control, communications, and dynamic games, including books on non-cooperative dynamic game theory, robust control, network security, wireless and communication networks, and stochastic networked control. He was the Editor-in-Chief of Automatica between 2004 and 2014, and is currently  editor of several book series. His current research interests include stochastic teams, games, and networks; distributed algorithms; security; and cyber-physical systems.
\end{IEEEbiographynophoto}
\vfill

\end{document}